\documentclass[sigconf]{acmart}

\AtBeginDocument{%
  \providecommand\BibTeX{{%
    \normalfont B\kern-0.5em{\scshape i\kern-0.25em b}\kern-0.8em\TeX}}}

\usepackage{marginnote}

\newcommand{\pageenlarge}[1]{\enlargethispage{#1\baselineskip}}

\AtBeginDocument{%
  \providecommand\BibTeX{{%
    \normalfont B\kern-0.5em{\scshape i\kern-0.25em b}\kern-0.8em\TeX}}}

\copyrightyear{2023}
\acmYear{2023}
\acmConference[Gen-IR@SIGIR2023]{The First Workshop on Generative Information Retrieval}{July 27, 2023}{Taipei, Taiwan}
\acmBooktitle{The First Workshop on Generative Information Retrieval (Gen-IR@SIGIR23), July 27, 2023, Taipei, Taiwan}
\acmDOI{}
\acmPrice{}
\acmISBN{}
\setcopyright{rightsretained}

\DeclareMathOperator*{\argmax}{arg\,max}
\DeclareMathOperator*{\argmaxs}{arg\,maxs}

\usepackage{amsmath}
\usepackage{natbib}
\usepackage{amsfonts}
\usepackage{graphicx}
\usepackage{subcaption}

\usepackage{hhline}
\usepackage{hyperref}
\usepackage{enumerate}
\usepackage{multirow}
\usepackage{array}
\usepackage[para]{footmisc}
\usepackage{marginnote}

\usepackage[show]{chato-notes}
\newcommand{\iadh}[1]{\textcolor{black}{#1}}
\newcommand{\craig}[1]{\textcolor{black}{#1}}
\newcommand{\cm}[1]{\textcolor{black}{#1}}
\newcommand{\craigws}[1]{\textcolor{black}{#1}}
\newcommand{\craigc}[1]{\textcolor{black}{#1}}
\newcommand{\xiao}[1]{\textcolor{black}{#1}}
\newcommand{\xiaow}[1]{\textcolor{black}{#1}}

\newcommand{\sm}[1]{\textcolor{black}{#1}}
\newcommand{\io}[1]{\textcolor{black}{#1}}

\newcommand{\xiaoJ}[1]{\textcolor{black}{#1}}
\newcommand{\craigj}[1]{\textcolor{black}{#1}}

\newcommand{\smm}[1]{\textcolor{black}{#1}}

\newcolumntype{P}[1]{>{\centering\arraybackslash}p{#1}}

\newcommand{\xa}[1]{\textcolor{black}{#1}}%blue
\newcommand{\smmm}[1]{\textcolor{black}{#1}}%red

\newcommand{\xf}[1]{\textcolor{black}{#1}}%blue

\author{Xiao Wang}

\affiliation{%
 \institution{University of Glasgow}
 \city{Glasgow}
 \state{Scotland}
 \country{United Kingdom}}
 \email{x.wang.8@research.gla.ac.uk}

\author{Sean MacAvaney, Craig Macdonald, Iadh Ounis}
\affiliation{\institution{University of Glasgow}
\city{Glasgow}
 \state{Scotland}
 \country{United Kingdom}}
\email{first.last@glasgow.ac.uk}

\begin{document}

% \title{Generative Query Reformulation}
\title{\xf{Generative Query Reformulation for Effective Adhoc Search}}

\begin{abstract}
\looseness -1 Performing automatic reformulations of a user's query is a popular paradigm used in information retrieval (IR) for improving effectiveness -- as exemplified by the pseudo-relevance feedback approaches, which expand the query in order to alleviate the vocabulary mismatch problem.
\xa{Recent advancements in generative language models have demonstrated their ability in generating responses that are relevant to a given prompt. In light of this success, we seek to study the capacity of such models to perform query reformulation and how they compare with long-standing query reformulation methods that use pseudo-relevance feedback.
In particular, we investigate two representative query reformulation frameworks, GenQR and GenPRF. GenQR directly reformulates the user's input query, while GenPRF provides additional context for the query by making use of pseudo-relevance feedback information. For each reformulation method, we leverage different techniques, including fine-tuning and direct prompting, to harness the knowledge of language models.} 
The reformulated queries produced by the generative models are demonstrated to markedly benefit the effectiveness of a state-of-the-art retrieval pipeline on four TREC test collections (varying from TREC 2004 Robust to the TREC 2019 Deep Learning). 
\xa{Furthermore, our results indicate that our studied generative models can outperform various statistical query expansion approaches while remaining comparable to other existing complex neural query reformulation models, with the added benefit of being simpler to implement.}
% \todo{SM: ``simpler to implement'' -- do we quantify this? Arguably the LLMs we use are super complicated to build -- we just don't need to because they already exist. Maybe tone this down somehow, e.g., ", despite its relative simplicity"?}
\end{abstract}

\maketitle

 \section{Introduction}
% \inote{outline:
% para1: 2 solutions: first solution : QE==> covers traditional QE methods and introduce our work is generating paraphrases;  and second solution is DE for solving vocabulary mismatch problem; ==>docT5query; 
% para2:pretrained NN models are good at various tasks: e.g. classification tasks: BERT re-rankers; BERTQE?
% and text generation task: for e.g. DE; but not for QR, thus in our work... i.e. T5QR
% para3: then we introduce the problem for vanillia T5 model and introduce QoPA CE loss. 
% }

% \inote{recheck abstract the introduction}
\looseness -1 Discrepancies between the way that users express their information needs and the content of relevant documents can cause such documents not to be retrieved or highly ranked. Several strands of research have tried to bridge the pervasive \xa{query-document mismatch problem in information retrieval \iadh{(IR)}, such as semantic or lexical mismatch,} namely dense embedding-based representations, document expansion and query expansion.

\looseness -1 Among the first strand, recent advances in pretrained neural network approaches have been shown to improve various text-processing tasks and to have the ability to learn the semantic meaning of the input text. \iadh{Indeed, various} approaches have demonstrated that BERT~\cite{devlin2019bert}-based deep neural ranking models are capable \sm{of capturing} \xa{the semantic and syntactic relationship between the texts, thus can mitigate the semantic mismatch between query and document}~\cite{nogueira2019multi,macavaney2019cedr,khattab2020colbert}. 
% to capture
% the latent traits of texts that are difficult for existing bag of words (BoW) models to apture~\cite{nogueira2019multi,macavaney2019cedr,khattab2020colbert}.% }\inote{this sentence needs reworded - dense retrieval is now more ubiquitious} a few recent advances in the use of dense retrieval mechanisms
However, apart from \xa{a thread of work in dense retrieval models ~\cite{karpukhin2020dense,khattab2020colbert,xiong2020approximate}, \xa{which struggle to handle long documents and suffer from limited interpretability}, many works have employed BERT models as neural {\em re-rankers}, in that they are applied to improve an initial ranking obtained from an inverted index using models such as BM25.}
\smm{In addition, industry has started performing BERT-based re-ranking at scale~\cite{Wang2021DomainSpecificPF}, demonstrating the value and applicability of this \xa{retrieve then re-rank} paradigm, \xa{which falls within the scope of this work.}} 
% \inote{I think we need to define scope here as not including dense.}

\looseness -1
\sm{To overcome the \xa{lexical} mismatch problem,} approaches such as query expansion and document expansion \iadh{are still promising},
in order to create a candidate set of documents with a sufficiently high recall to enable a \xa{neural re-ranker}
% BERT model
to identify and up-rank relevant documents. Document expansion augments each document with additional information, for instance, additional terms selected from \iadh{a} corpus~\cite{billerbeck2005document,dai2020context} or predicted queries generated by a natural language generation model from the original document. Indeed, the so-called doc2query approach initially \xa{expands} each document by generating multiple predicted queries using \xa{generative language models~\cite{nogueira2019document, nogueira2019doc2query}.}
% cite,cite. sequence-to-sequence transformer model~\cite{nogueira2019document} and later using the T5 model~\cite{nogueira2019doc2query}. 
This is effective but comes with the significant \sm{upfront} cost of applying a \xiao{heavy} model to every document before indexing.

\looseness -1 On the other hand, query expansion typically describes approaches that modify the users' query with the aim to improve the retrieval effectiveness. By adding additional terms selected from the pseudo-relevant set of returned documents~\cite{croft2010search} in response to the initial query or by expanding the original query with lexical-level~\cite{zukerman2002lexical} or phrase-level~\cite{riezler-etal-2007-statistical} paraphrases, the distance between the (reformulated) query and the relevant document(s) is \craig{reduced}. \craigc{Similarly, \xa{in the more recent}
\xiaoJ{CEQE}~\cite{naseri2021ceqe} method, expansion terms from the pseudo-relevant documents are identified in the BERT embedding space.} These approaches are compatible with neural re-rankers, in that a refined ranking \xiao{list} obtained using query expansion can improve the effectiveness of a BERT\iadh{-based} re-ranking \iadh{model}~\cite{nogueira2019multi,macavaney2019cedr,khattab2020colbert}. \smmm{In addition,}  neural \iadh{pseudo-relevance feedback (PRF)} models have been proposed, such as  NPRF~\cite{li2018nprf} and BERT-QE~\cite{zheng2020bert}.
% \todo{SM: where do ColBERT-PRF, ANCE-PRF, etc. come in?}
\iadh{However,} these models are limited in their functionality, in that the additional relevance signal obtained from the pseudo-relevant set is only used to re-rank the candidate set of documents, rather than creating a higher recall candidate set by re-executing a reformulated query on the inverted index. 
\xa{Additionally, several dense-PRF methods, such as ColBERT-PRF~\cite{wang2021pseudo,wang2022colbert} and ANCE-PRF~\cite{yu2021improving}, have been proposed to make used of the BERT pretrained knowledge to improve query representation in dense retrieval models using PRF knowledge.}
Instead, \iadh{our work in this paper} aims to \sm{investigate the potential of} \xa{generative} neural \io{models} to produce a refined candidate set of documents by \xiao{generating} a refined query reformulation \smmm{for a sparse retrieval approach.}

\looseness -1 Indeed, motivated by the fact that the same information need can be formulated using different natural language expressions, in this \io{paper}, we focus on expanding the original query by generating {\em paraphrases} that share the same information need, to improve the retrieval effectiveness. 
We cast the query reformulation task as a \smmm{text generation} task. 
This allows us to be able to test whether the knowledge encapsulated by {\em pretrained} \smmm{text generation}
\xa{models} --- \xa{such as T5~\cite{raffel2020exploring} and a more advanced instruction-tuned T5 variant called FLAN-T5~\cite{wei2021finetuned} ---} can be exploited for query reformulation. 
\xa{We explore two possible generative query reformulation frameworks, GenQR and GenPRF.}
\sm{The first produces reformulations using only the user's query text itself (GenQR). The second approach makes use of pseudo-relevant documents as \iadh{contextual} information \xiao{(GenPRF)}, to guide the reformulation process.}
% \xa{Each query reformulation framework can be instantiated leveraging T5, where we inject the knowledge of query reformulation task by further fine-tuning the T5 model, and the FLAN-T5, where we unlock the knowledge of the FLAN-T5 model via prompting.  }
\smmm{For T5, we explore techniques to fine-tune the model for each query reformulation task.}
% As we will show, our T5-based models can be instantiated to reformulate \xiao{the query} using only the user's query as input (which we call T5QR), or to additionally make use of pseudo-relevant documents as \iadh{contextual} information \xiao{(T5PRF)}, to guide the reformulation process.
% (T5PRF)
% \xa{The overall architecture of our \iadh{proposed} framework is shown in Figure~\ref{fig1}. }
\cm{One challenge \smmm{we found} \iadh{when} \craigc{fine-tuning} the T5 model for query reformulation is that there \iadh{are} no gold-standard labelled query \craigc{reformulations}. To circumvent the lack of ground-truth data, we
% \iadh{propose to use}
\sm{investigate the use of} weakly supervised query pairs to fine-tune the T5 model, instead of human annotated query pairs. In \craig{particular}, we leverage \xiao{three} different \craigws{filters that improve the} weakly supervised dataset \craigc{to reduce the noise present in the query pairs and thus reduce the training time of \xa{GenQR and GenPRF} when injecting the task-specific knowledge into the T5 model.}}
\smmm{Meanwhile, for FLAN-T5, we explore prompting techniques in attempt to leverage the patterns observed in the pre-training process.}
% \todo{SM: is there anything more we can say here about the challenges we encountered for FLAN-T5? E.g., it's highly sensitive to the prompts? XIAO: in the discussion section, no need to repeat here?}

\looseness -1 In summary, our work makes four contributions: \xa{(1) we propose two generative query reformulation frameworks, namely GenQR and GenPRF; (2) we make use of two generative models, a pretrained language model, T5, and an instruction-tuned language model, FLAN-T5, to reformulate input queries;}
% (1) we make use of a pretrained T5 model to paraphrase (reformulate) input queries; \cm{(2) we \iadh{propose} different filters for enhancing the training of  \iadh{the} T5 model through weak supervision;} 
(3) we make use of the \iadh{contextual} \xa{pseudo-relevance feedback} information as input for \iadh{the} \xa{generative model;}
% T5 model;
(4) we demonstrate the effectiveness of the generated query reformulations \iadh{in comparison to several existing baselines} within the \xa{prevalent retrieve then re-rank pipeline;} 
% \inote{still hold?} state-of-art retrieval pipeline; 
\sm{and} \xiaow{(5) we investigate the effectiveness of the neural re-ranker when combined with \xa{GenQR and GenPRF} results. }%\inote{check this sentence}
% \cm{More generally, our work investigates how to \io{effectively} use BERT-based neural re-rankers when  \inote{generating?} expanded queries.}

%\todo[craig]{we need another paragraph here about combining with DNN re-rankers}

% The remainder of this paper is organised as follows: Section~\ref{sec:related_works} describes related work about query reformulation, while Section~\ref{sec:T5QR} presents our \xa{proposed generative query reformulation frameworks.} Research questions and the used experimental setup are detailed in Sections~\ref{sec:RQ} \& ~\ref{sec:experimental_setup}, \io{respectively}. Next, we \io{report and} discuss our experimental results in Section~\ref{sec:results}. Finally, we summarise our findings and provide future work directions in Section~\ref{sec:conclusion}.\todo{SM: if we need space, we can drop this paragraph; the paper is organized conventionally, so doesn't really need explanation.}

\section{Related Work}\label{sec:related_works}

%We describe two aspects of related work, namely a review of the query reformulation models for adhoc retrieval and query performance predictors.

% \todo[]{related works:

% 1. related works for query reformulation; rewirte it.

% ==>traditional QR methods: 1 paragraph: QE; PRF, explicit QE. click-logs based query-level paraphrase; SMT method paraphrase;
% 2. neural QR mdoels
% 2.1. neural networks for QR methods:  seq2seq model; RL model; however, train from scratch. 
% \inote{The most close work to us is that Want et al. NOT GONNA COMPARE WITH DRQRmodel}

% 2.2. large pretrained model for QR: conversational QR; T5; GPT2, however, lack of human annotated training data, not yet used for ad-hoc. 

% 3. Query performance prediction }

% \subsection{Query Reformulation Models}
\pageenlarge{2} \looseness -1 Many approaches have been proposed to address the vocabulary mismatch problem in information retrieval \iadh{(IR)}, typically by modifying the user's original query. Popular
pseudo-relevance feedback \io{approaches include models} such as the Rocchio algorithm~\cite{croft2010search}, the RM3~\cite{abdul2004umass} relevance language model, or the DFR Bo1 query expansion model~\cite{amati2003probability}. All of these approaches identify candidate terms from the top returned results in response to the original query, which are \iadh{then} added to the query with weights representing their importance in conveying the query, and resulting in a larger {\em expanded} query. The expanded query consists typically of \cm{a weighted bag of} related terms, but does not constitute a well-formed sentence. 
Other query expansion approaches \iadh{involved} \io{reformulations} based on \iadh{the use of thesauri and dictionaries}~\cite{jing1994association}
% \inote{ADD citations; e.g. see Croft book} 
or \iadh{the} analysis of similar queries in query logs~\cite{craswell2007random,jones2006generating}. 
% \inote{we need to add a description of a QE using WordNet for example}
For instance, Jones et al.~\cite{jones2006generating} replaced phrases within queries with those in similar queries \xiao{and Zukerman et al. used an iterative process to identify similar phrases to a query based on WordNet~\cite{zukerman2002lexical}}. \xiao{Moreover, word embeddings have also been explored to perform query expansion~\cite{diaz2016query,kuzi2016query}. } 
However, these query expansion models rely on the information extracted from the retrieved documents or external resources (e.g. click data or WordNet) to enhance the representation of the query.
% \xiaoJ{More recently, }
% \inote{this is different from what we said in the introduction}
\xiao{In contrast,  our approach focuses on generating \craig{whole refined textual queries} using a text generation model.}

\looseness -1 Recently, neural text generation models have been \xiao{gradually }
% extensively\inote{started to be}
 explored for query reformulation.  For instance, in~\cite{shankar2019legal}, a RNN-based encoder-decoder neural model \craig{was} employed to reformulate an input query, where the model is trained to \xiao{refine the noisy input text}.
%  learn a better representation of the input text.
%  \inote{legal comes from nowhere; woudl this model generalise to adhoc?}
%  \inote{only one example for "extensive"}.
\xiaow{\craigc{A} sequence-to-sequence \craigc{model} with \craigc{an} attention and copy mechanism \craigc{has also been} used to generate the suggested query based on previous query \io{sessions}~\cite{dehghani2017learning}.}
 Moreover, reinforcement learning-based neural network models have been explored to \xiao{select relevant terms to expand the users' query~\cite{nogueira2017task,montazeralghaem2020reinforcement}.}
In addition, Wang et al.~\cite{wang2020deep}, 
% whose work is the most similar to our present efforts,
% \todo{SM: is it still really the most similar to ours now, given the current landscape?}
proposed to 
% \inote{we need to say how it goes further than nogueria}  
incorporate the query performance prediction signal of the query into the process of query reformulation without reliance on the initial retrieval phase. However, all of these existing models require training from scratch, and hence are not able to make use of pretrained transformer models that have recently become popular. 
In contrast, we make direct use of the knowledge encapsulated by a {\em pretrained} sequence-to-sequence transformer model for tackling the adhoc query reformulation task. %\inote{strong gap? yes good i think}
% \todo[craig]{Xiao related work must be in PAST TENSE}

\looseness -1 Transfer learning, \xiao{where the knowledge stored for addressing one task can be transferred to address other related tasks,} 
% \inote{is the definition of transfer learning clear, how this links to T5} 
provides flexible ways to leverage the knowledge encapsulated in large pretrained models, such as T5~\cite{raffel2020exploring} and BERT~\cite{devlin2019bert}. \craigc{As a result,}  many researchers have turned to \iadh{making} use of such knowledge \xiao{to address various downstream tasks.} 
% to learn the contextual meaning of the input text.
\smmm{Text generation models have been extensively explored for the task of \textit{document} expansion (e.g.,~\cite{nogueira2019document,doct5query,DBLP:conf/ecir/GospodinovMM23}), where generation models are trained to produce relevant tokens that can be appended to documents before indexing. However, these approaches are limited by the considerable compute required to run the expensive generation process for all documents in the collection.}
% \inote{for re-ranking, monoT5.}
\xiaoJ{In addition, the pretrained knowledge of the BERT model has been explicitly employed to perform neural query expansion for adhoc search. For instance, Neural PRF~\cite{li2018nprf} and BERT-QE~\cite{zheng2020bert} employ BERT model to select the chunks of the relevant text together with the original query text to score a given document. CEQE~\cite{naseri2021ceqe} selects expansion terms in the BERT embedding space using three methods, namely CEQE-Max, CEQE-Mul and CEQE-Centroid, then adds the expansion terms to the original query terms to form a new query. Thus, in CEQE, terms in the feedback documents that are contextually similar to the original query are highly weighted in the reformulated query.} \xa{Moreover, dense-PRF methods,  such as ColBERT-PRF~\cite{wang2021pseudo,wang2022colbert}, ANCE-PRF~\cite{yu2021improving} and Vector-PRF~\cite{li2021pseudo}, make use of the BERT knowledge and the PRF information to augment the query representations for dense retrieval. In contrast, instead of working around BERT which has limited interpretability, we opt for a generative method for query reformulation and concentrate on well-optimised sparse retrieval.}
% are benefited from the BERT}
% \todo{SM: again, ColBERT-PRF, ANCE-PRF}

% \inote{add a paragraph about the prompt, and LLM, flan-t5}

\pageenlarge{2} \xiaoJ{On the other hand, for OpenQA task, Mao et al.~\cite{mao2020generation} used \craigj{a} pretrained language model as \craigj{a} generator to produce background context information for an input question -- this generated context together, with the input question. formed an augmented query  task. However, this framework is not applicable for addressing adhoc search task due to the lack of \craigj{a} training dataset, e.g. query and ground truth answer training pairs, \craigj{while the produced augmented queries are long in length which} can result high query latency during retrieval.}
%however, 
More \craig{similar} to our work, Mass et al.~\cite{mass2020unsupervised} fine-tuned a GPT-2~\cite{mass2020unsupervised} model using FAQ (Frequently Asked Questions) pairs to generate \iadh{the} predicted questions given an input query. In addition, T5~\cite{raffel2020exploring} or GPT2~\cite{radford2019language} \iadh{have} been employed as neural transfer reformulation \iadh{models} to reformulate the query based on the recent querying history in conversational search~\cite{lin2020conversational,lin2020query,yu2020few}. \xiao{Furthermore, \io{\citet{lin2020conversational}} \cm{compared} pretrained models using an encoder-decoder architecture (T5 model) and a GPT2 (decoder only) model for  conversational question reformulation, and concluded that T5 is more effective.}

% In particular, for fine-tuning a trained language model, we leverage weak supervision methods to generate the necessary training query pairs, which are then used to fine-tune the existing pretrained T5 model for the query reformulation task.

\smmm{While prior work has used smaller open-source models, there has been a recent shift to very large, proprietary, closed-source text generation models, such as OpenAI's {\em text-davinci-003} with 175 billion parameters.
Concurrently with this }\xa{work, \citet{wang2023query2doc} proposed query2doc method, which expands the query with generated pseudo-documents from {\em text-davinci-003}. In addition, \citet{iain2023grf} also prompt {\em text-davinci-003} for query expansion using various prompting techniques. \smmm{Unlike these works, we provide a} comparison between the effectiveness of fine-tuning and prompting methods for a particular task. \smmm{Further, we explore multiple frameworks for query reformulation,} namely GenQR and GenPRF, \smmm{and investigate} how to more effectively employ the learned knowledge from the large neural models. In addition, it is well-established that large language models (LLMs) with parameters exceeding 100 billion have superior performance on various tasks~\cite{brown2020language,chowdhery2022palm,weiemergent}. However, \smmm{using these} large-scale models is often prohibitively expensive. Therefore, our work explores the ability for using small models with parameters of less than 10 billion} for which inference can be performed on consumer GPUs.
% (specifically, {\em t5-base} with 220 million parameters and {\em flan-t5-xxl} with 3 billion parameters) 
% for reasonable query reformulation performance.}
% In particular, for fine-tuning a trained language model, we leverage weak supervision methods to generate the necessary training query pairs, which are then used to fine-tune the existing pretrained T5 model for the query reformulation task.}

% \inote{Here add the LLM for query expansion applications, query2doc. However}
% However, the application of a large pretrained model has not been sufficiently explored for the \xiaoJ{query reformulation generation}
% reformulation of queries 
% in adhoc search, \cm{\iadh{possibly} \craig{due} to the lack of human annotated ground-truth query reformulations. 
% In this work, we firstly leverage weak supervision methods to generate the \iadh{necessary} training query pairs, which are then used to fine-tune \craigc{the existing pretrained} T5 model for \iadh{the} query reformulation task.

% \section{A T5-based Model for Query Reformulation}
\section{Generative Query Reformulation}\label{sec:T5QR}
% \inote{rewrite here:} 

\xa{In this section, we first define the query reformulation problem \xiaow{and detail the GenQR model with knowledge injection and prompting} in Section~\ref{problemStatement}.  
\xiaow{Then we describe our GenPRF model in Section~\ref{context_inputs}. Finally, the weak supervision \craigws{data process used to \xa{fine-tune} the proposed GenQR and GenPRF when performing the knowledge injection method, and how it is filtered for quality is} introduced in Section~\ref{weak_supervision}.}}

% \craig{\iadh{The use} of pseudo-relevance feedback as context input  is detailed in Section~\ref{context_inputs}. Finally, the weak supervision \craigws{data process used to train T5QR, and how it is filtered for quality is} introduced in Section~\ref{weak_supervision}.}

% describe our proposed T5QR model with \iadh{a} query performance \iadh{prediction-based} regularised cross-entropy loss.

\subsection{GenQR}\label{problemStatement} % Query Reformulation Problem Definition
% \inote{introduce PRF process; then follow with the \textbf{knowledge injection} and the \textbf{prompt engineering} paras}
Query reformulation can generally be expressed as a process $\mathcal{P}$ that takes the user's initial query $q^0$ and converts it into a new representation $q^r$, with the aim of improving retrieval effectiveness: 
% \inote{why you have ... in eq (1)? you need to say what it means}:
\begin{equation}
\small
q^r = \mathcal{P}(q^0, ...)
\end{equation}%\inote{what is $q\prime$}
\xiao{\noindent where "..." denotes the optional information that may \iadh{be used by} a query reformulation process $\mathcal{P}$. }
For instance, a query rewriting process might rely on the text of the query alone:
\begin{equation}\small
q^r = \mathcal{P}_{\xiao{QR} }(q^0) \label{eq:queryreform}
\end{equation}

On the other hand, many query expansion approaches obtain a reformulated query through the application of a classical pseudo-relevance feedback mechanism -- such as Rocchio~\cite{croft2010search} or RM3~\cite{abdul2004umass} -- which \iadh{makes} use of query terms occuring in the top-ranked documents \iadh{returned} for the original query $q^0$, as follows:
\begin{equation}\small
q^r= \mathcal{P}_{PRF}(q^0, {R}_{K}(q^0)), \label{eq:PRF}
\end{equation}
where ${R}_{K}(q^0)$ is a ranking of $K$ documents \craig{obtained} using $q^0$ \craigc{and $q\prime$ takes the form of a weighted set of terms.}

% \craig{In both scenarios \inote{we need names for these two scenarios, then put their names here} cases, the final ranking of documents is obtained by obtaining the final ranking of documents using $q^\prime$, as $R(q^\prime)$.}

% \pageenlarge{2} 
\looseness -1 \xiao{In both scenarios, $\mathcal{P}_{QR}$ and $\mathcal{P}_{PRF}$ can be \iadh{seen} as methods \iadh{that add} candidate terms or phrases to \xiao{rewrite} or expand the original query \craigc{(with \io{the} optional weighting of \io{such} terms)}.} Instead, we rely on a T5 text generation model for $\mathcal{P}$, which can refine the initial query $q^0$ by generating paraphrases of the original query. In addition, to ensure that the paraphrases of the original query are on-topic \iadh{in relation to} the initial query, \io{we also study an} approach that can also encapsulate pseudo-relevance information as context during the reformulation process. Thus, \xa{we study two generative query reformulation frameworks:} \iadh{\io{one} that \io{takes} the} form of a general query reformulation process (i.e. GenQR), \xa{along with a pseudo-relevance feedback-based query reformulation (i.e. GenPRF)}. \xa{For each query reformulation paradigm, we introduce two methods, namely fine-tuning and prompting, to leverage the pretrained knowledge of large language models for query reformulation.}

% (cf. Equation~\eqref{eq:queryreform}), \io{along with} a pseudo-relevance feedback-based query reformulation \io{approach} (cf. Equation~\eqref{eq:PRF}). 

% In addition, relevant information of the pseudo-relevant feedback document set can act as the contextual background for our neural query reformulation model. 

% \begin{equation}
% q\prime = \mathcal{P}_{Reg-T5QR}(q^0, ...)
% % \label{equ:T5QR}.
% \end{equation}

% \inote{check}

%  Figure~\ref{fig1} shows the overall framework of the neural query reformulation model in IR. We first fine-tune a pretrained T5 model using our proposed QPP-based regularised loss. The obtained reformulated query concatenate to the original query as a new query to performance retrieval task.  Each pair of training queries consists of a source query and a target query, where the source query is prepended with the word ``paraphrase'' and suffixed by the end-of-sequence token ``$\langle /s \rangle$''.

% At inference time, the obtained T5QR model can be used to generate a query reformulation $q\prime$ in response to $q^0$.
% The output token is sampled from the $Reg-T5QR(\hat{\theta})$ model, predicted output distribution one at a time until the generation of the special end-of-sequence token. 
% For instance, given the original source query ``paraphrase: do goldfish grow $\langle /s \rangle$'', the T5QR model might generate the reformulated query ``how long does goldfish grow $\langle /s \rangle$''.

%  \subsection{T5QR}~\label{Traditional_T5QR}

% \todo[craig]{make a section here about T5QR. Include the traditional loss function}

\def\reprompt{\textsf{``refine''}}
\def\coprompt{\textsf{``context:''}}
\def\eos{\textsf{``</s>''}}

\textbf{\xa{Fine-tuning:}}
Formally, our \xa{generative} query reformulation process \xa{via injecting the task-specific knowledge into a pretrained \smmm{text generation} model (a pretrained T5 model) is denoted as $\mathcal{P}_{T5QR}$. In particular, the $\mathcal{P}_{T5QR}$ takes the initial query $q^0$ (in the form of a sequence of text \iadh{and} optionally other information) and produces refined queries. }
% \xa{Figure~\ref{fig1} illustrates the working stages of our proposed T5QR model.}
% , \iadh{denoted by} $\mathcal{P}_{GenQR}$, can be described as a Text-to-Text framework \iadh{that} takes as \craig{input the initial query $q^0$ (in the form of a sequence of text \iadh{and} optionally other information)} and produces refined queries. 
When conditioned only on the input query $q^0=q^0_1, ... q^0_{|q^0|}$, \craigc{a query paraphrase can be} generated by applying a fine-tuned text generation function, $T5()$, as follows:
\begin{equation}\small
\texttt{T5}(\reprompt, q^0_1, ... q^0_{|q^0|},\eos)\nonumber
\end{equation}
where in the input sequence, the input query is prepended with a special prompt token $\reprompt$ indicating to T5 that it should reformulate the input and \iadh{it is} suffixed by the \iadh{special} end-of-sequence token $\eos$. 

% \begin{figure}[tb]
% \includegraphics[width=\linewidth]{fig/T5QR_workshop.png}%  
% \caption{The proposed T5QR model for fine-tuning a pretrained model for the adhoc query reformulation task.}\label{fig1}  
% \end{figure}

% %%%%%%%%%%%%%%%%%%%%%%hide for generative workshop, maybe put in the appendix%%%%%%%%%%%%%%%%%%%%%%
% \looseness -1 \xa{After fine-tuning, we obtain our $\mathcal{P}_{T5QR}$, }\craigc{the final output of $\mathcal{P}_{T5QR}(q^0)$ is \craigc{obtained by combining the output sequences of $N$ applications of the T5 model, and weighting the terms from each sequence by the joint likelihood of the sequence, as follows:}}
% %as the weighted combination of terms:}
% \begin{equation}
%     \mathcal{P}_{T5QR}(q^0) = w_{q^{r1}}\cdot \left[ q^{r1}_1,... q^{r1}_{|q^{r1}|}\right]+...+ w_{q^{rN}}\cdot \left[ q^{rN}_1,... q^{rN}_{|q^{rN}|}\right],\label{eqn:PT5QR}
% \end{equation}
% \xiaow{where $w_{q^r}$ denotes the joint likelihood of reformulation (or paraphrase of the input query) $q^r$ -- i.e.\ $w_{q^r} = P\left(q^r_1, q^r_2,..., q^r_{|q^r|}\right)$} -- and $N$ is the number of predicted output sequences used to form a query reformulation in response to the original query $q^0$. \craigc{By generating and combining $N$ paraphrases of the original query generated by T5, important terms are more likely to receive higher weights -- indeed, a similar repeated application of T5 is used by docT5query~\cite{nogueira2019doc2query}.}.

%%%%%%%%%%%%%%%%%%%%%%hide for generative workshop, maybe put in the appendix%%%%%%%%%%%%%%%%%%%%%%

\textbf{\xa{Prompting:}}
% \inote{how to formulate the FLAN version}
% $\mathcal{P}_{\texttt{FlanQR}}$
\xa{Instead of injecting task-specific knowledge via fine-tuning a language model, FLAN-T5 employs an instruction-tuning approach to improve the ability of a language model for various tasks. However, the capability of FLAN-T5 for performing query reformulation task is still unclear. Thus, besides fine-tuning a T5 model, we also investigated employing the FLAN-T5 model for reformulating the initial user query and we denote this query reformulation process as $\mathcal{P}_{FlanQR}$. More specifically, we design a query reformulation task aware prompt and combine it with the original query $q^0$ as the input for $\mathcal{P}_{FlanQR}$. The input template for $\mathcal{P}_{FlanQR}$ is as follows,}
\begin{equation}\small
    \texttt{FlanQR}({<\texttt{Prompt}>}, q^0_1, ... q^0_{|q^0|} ).
\end{equation}

However, while \io{the user's} input queries may be short and \io{may not} provide sufficient evidence to interpret the meaning of the input query. \xa{We hypothesise that the additional pseudo-relevant context information can aid the query reformulation process. In the next section, we show how T5QR and FlanQR can be formulated as pseudo-relevance feedback mechanisms.}
% following Equation~\eqref{eq:PRF}.
% \inote{put in the appendix?}

\subsection{GenPRF}\label{context_inputs}

% \pageenlarge{2} 
\looseness -1 To reinforce the capability of the query reformulation model to interpret the meaning of the input query, an additional context that further explains the query statement can be desirable. Motivated by the utility of the top returned documents in pseudo-relevance query expansion models (such as Rocchio~\cite{croft2010search}, Bo1~\cite{amati2002probabilistic} and RM3~\cite{abdul2004umass}), we leverage the top returned documents for the original query and use them as the pseudo-relevance contextual information for the query. 

\textbf{\xa{Contextual Fine-tuning:}} 
The query together with its contextual information is taken as the input to \xiaow{fine-tune a pretrained T5 model}:
\begin{equation}
\small
\texttt{T5PRF}(\reprompt, q^0_1, ... q^0_{|q^0|},\coprompt, context(R) ,\eos)  \label{CAequ}
\end{equation}
\noindent \looseness -1 \craig{where (similar to $\reprompt$) $\coprompt$ is a \craig{prompt} token denoting the start of the context passage(s); $context(R)$ identifies important passage(s) of text from the pseudo-relevance feedback document set $R$.} \xa{Thereafter, we instantiate this approach under the GenPRF framework as T5PRF.}
% The final reformulated query for T5PRF is obtained by combining the weighted combinations of $N$ invocations of Equation~\eqref{CAequ}, in a similar manner as for T5QR in Equation~\eqref{eqn:PT5QR}\inote{check this eqn nbr}.

% \pageenlarge{2}
\textbf{\xa{Contextual Prompting:}}
% \inote{what the input template for this method?}
Additionally to T5PRF, we also instantiate the GenPRF as FlanPRF, where we direct prompt the FLAN-T5 model using a task-aware prompt, the initial query $q^0$ as well as the top-ranked pseudo-relevance feedback information $context(R)$ as input. The input template for FlanPRF can be expressed as follows,
\begin{equation}
\small
    \texttt{FlanPRF}({<\texttt{Prompt}>}, q^0_1, ... q^0_{|q^0|}, context(R) ).
\end{equation}

\textbf{\xa{Types of Contextual Information}}
\io{\xa{Rather than use whole documents,} we use passages as \craigc{pseudo-relevance} context information. Indeed, two} considerations prevent entire feedback documents being used as context information. \smmm{Indeed, \io{the} compute and memory requirements of the transformer architecture rise exponentially as sequence length increases, so lengthy documents can easily exceed available memory.} Moreover, long documents may not be \craigc{wholly concerned with} the topic of the query, potentially misleading \xa{our PRF models} in understanding the meaning of the input query. 
% \xiao{To address these concerns,} we leverage three different approaches to obtain \iadh{pseudo-relevance} information, as shown in Figure~\ref{fig3}.
The whole process can be described as \iadh{follows}: Firstly, for a given initial query $q0$, we obtain the first $K$ documents returned by the retrieval model,  i.e. $R_K(q0)$; 
\io{Secondly}, a sliding window \iadh{with size} $w$ is used to break each document into passages with an overlap of $w/2$ words between \xiao{neighbouring}
passages -- we denote the text passages obtained from document $d$ as $passages(d)$; Thirdly, a ranking model is employed to assign a score for each passage, measuring the relevance of the content of the passage to the query; finally, we apply one of three \craig{selection mechanisms to obtain} the \xiao{context} information for \xiao{the T5PRF model}, \io{namely the FirstP, TopP and MaxP selection approaches}.

The intuition behind \xiao{FirstP} is \iadh{that} the leading paragraph in a long document often contains \iadh{the} \xa{main} gist \iadh{of the} \xa{document}. Hence,  we pick the $M$ highest scoring among the first passages of all documents in $R$, as follows:
\begin{equation}\small
context_{FirstP}(R, q^0) = \argmaxs^M_{p\in passages(d)[0],\, \forall d \in R}s(p,q^0)
\label{eq:PRF-FirstP}
\end{equation}
\looseness -1 where $passages(d)[0]$ denotes the first passage obtained from document $d$, and $s(p,q^0)$ denotes the relevance score given by a ranking model between a passage $p$ and the original query. $\argmaxs^M()$ is an extension of $\argmax$, \io{which} returns the top $M$ scoring items.

Next, rather than obtaining the first passages, we look to identify the most relevant passages from the feedback set $R$, inspired by the MaxPassage ranking approach~\cite{dai2019deeper,liu2002passage}. In particular, we formulate two selection mechanisms, namely \xiao{TopP and MaxP}, \craigc{which slightly differ in how they obtain the highest $M$ scoring passages from $R$.} \xiao{TopP} takes the $M$ highest scoring passages across all of the feedback set, while \xiao{MaxP} selects the highest scoring passages from each document and then \io{selects} the highest $M$ among \io{these}:
\begin{eqnarray}\small
context_{TopP}(R, q^0) = \argmaxs^M_{p \in passages(d),\, d \in R} s(p, q^0)  \\
% \label{eq:PRF-TopP}
% \end{equation}
%\begin{equation}
context_{MaxP}(R, q^0) =  \argmaxs^M_{ d \in R } \argmax_{p \in passages(d) }   s(p, q^0)
\label{eq:PRF-TopP}
\end{eqnarray}

% \begin{figure}[tb]
% \includegraphics[width=.7\linewidth]{fig/Context.png}  
% \caption{Different \io{types} of contextual information 
% % \inote{a better figure? think}
% }
%   \label{fig3}
% \end{figure}

% Assuming $M=1$,  Figure~\ref{fig3} illustrates the different feedback selection mechanisms. In particular, \xiao{FirstP} selects only a first passage, while  \xiao{MaxP and TopP} identify different passages within the feedback set.

\subsection{Weakly Supervised Query Pairs and Filters}\label{weak_supervision}
% \looseness -1 \xa{When implementing the T5QR and T5PRF models, we need to fine-tune T5 model.} 
In order to fine-tune T5
% which is a sequence-to-sequence \iadh{architecture-based} model 
\xa{for the T5QR and T5PRF models,} we need a large number of training pairs that contain 
\xa{a} labelling signal. However, for the adhoc query reformulation task, there is no readily-available large-scale labelled dataset of query reformulations. Hence, to circumvent this shortage, we leverage existing test collections to generate the required weak supervision signal for T5.

In particular, to allow T5 to learn how to generate a paraphrase in response to an input query, the training pairs should convey the same information need. \io{In other words,} they should be semantically similar to each other. Following the work of Zerveas et al.~\cite{zerveasbrown}, we first identify pairs of queries that share at least one relevant document from a test collection.
This forms our initial pool. Our main underlying assumption is that if a document is labelled as relevant for multiple queries, such queries are assumed to convey the same information need~\cite{wang2020deep,zerveasbrown}. 
However, a clear risk when directly using the initial training pool is that noise can arise -- for instance, the content of a document might address multiple different topics, or the queries pertain to different information needs. Hence, to reduce this noise, we introduce three filters that can be applied individually -- or in combination, -- in order to further improve the quality of the initial training pool, \craigc{as well as reduce training time}.

\xf{More specifically, we first 
identify all pairs of queries that are associated to the same relevant document. 
% generate all possible combinations of queries, documents and relevance labels, then filter the tuples where the relevance label is ``relevant''.
% Next, we group the tuples with the same document and 
Then we extract the pairs of queries from these tuples as the initial pool $\mathcal{M}$. }
Within each query pair $\langle q_x, q_y\rangle$ in $\mathcal{M}$, each query consists of a sequence of words, i.e. \xiao{$q = (q_{1}, q_{2},\cdots q_{\vert q\vert})$}. To refine $\mathcal{M}$, we apply filters to reduce the number of query pairs, or refine those query pairs to provide more useful reformulations for fine-tuning T5, i.e.\ $\mathcal{M'} = \mathcal{W}(\mathcal{M})$. We now introduce our \io{three} proposed filters.

\subsubsection{Overlap \craigws{Filter}}
This filter assumes that only query pairs for which there is a marked overlap in the retrieved documents for both queries are suitable training examples. A similar approach was previously described in~\cite{cronen2004framework} for identifying \iadh{a} query drift in query expansion. Similarly, Mass et al.~\cite{mass2020unsupervised} employed this method to identify \iadh{a} potential topical drift when generating paraphrases of FAQ questions. In particular, we use an overlap-based filter -- denoted as $\mathcal{W}_O$ -- for selecting high quality query pairs from the initial pool $\mathcal{M}$. 
For a pair of queries $\langle q_x, q_y \rangle \in \mathcal{M}$, we issue $q_x$ and $q_y$ against the corpus index, respectively, and check the number of shared documents in the returned top-K result lists. 
The overlap of the result lists is defined as the cardinality of the intersection of the top $K$ documents of the two \iadh{result lists} for $q_x$ and $q_y$: 
\begin{equation}\small
    O(\langle q_x,q_y\rangle) = \vert R_K(q_x)\cap R_K(q_y) \vert  \label{equ:overlap}
\end{equation}
Hence, the overlap weak supervision filter is defined as: 
\begin{equation}\small
    \mathcal{W}_O(\mathcal{M})= \{ \langle q_x,q_y\rangle \in \mathcal{M} \land O(\langle q_x,q_y\rangle) \geq \delta_{O} \},
\end{equation}
\looseness -1 where $\delta_{O}$ defines a threshold on the minimum overlap -- for higher values of $\delta_{O}$, there is less chance of topical drift between queries.

  \subsubsection{Effectiveness Filter}

\looseness -1 The aim of fine-tuning T5 for the query reformulation task is to {\em improve} the retrieval effectiveness in response to an input query. Hence, when selecting training pairs, the retrieval performance should be taken into consideration, such that the target query $q_y$ in the training dataset \iadh{has} a better quality than the source query $q_x$ (in other words, the target query leads to a better retrieval performance than the source query). 
Therefore, we introduce the Effectiveness Filter, denoted as $\mathcal{W}_E$, to improve the quality of the initial pool.
Given a query pair $\langle q_x, q_y\rangle$, we measure their relative effectiveness, with reference to relevance assessments $L$. Let $M(q)$ \io{denote} the effectiveness of ranking $R(q)$ for a particular effectiveness metric, such as reciprocal rank (RR) or discounted cumulative gain (DCG). \craigc{Then}, $\mathcal{W}_{E}$ filters \iadh{the} query pairs based on their relative effectiveness, as follows: 
\begin{equation}\small
    \mathcal{W}_E(\mathcal{M})= \{ \langle q_x,q_y\rangle \in \mathcal{M} \land (M(q_y) - M(q_x) > \delta_{E} \}
\end{equation}%is a threshold on
\looseness -1 where $\delta_{E}$ \craigc{defines} the \craig{required minimum} positive change in $M$.%effectiveness.

\subsubsection{Stopwords Filter}
Stopwords are function or grammatical words, e.g.\ ``is'', \iadh{or} ``and'', which have high term frequencies but contribute very little to enhancing the retrieval performance~\cite{roy2019selecting}. However, many query pairs may consist of syntactical variations of the same query, such as \craig{stopwords that have changed order, or have been substituted with other stopwords.}
In order to help the pretrained T5 model focus on generating terms that can identify relevant documents\footnote{\craigc{Indeed, in our experimental setup, stopwords are removed at retrieval time, so do not contribute to effectiveness.}}, we propose the Stopwords Filter -- denoted as $\mathcal{W}_{S}$ -- which refines \iadh{the} initial pool conditioned on the presence of the common words or stopwords.
Given a pair of queries $\langle q_x,q_y\rangle$, to aid the T5 model \iadh{producing} non-generic words given any natural language query, the stopwords of the target query $q_y$ are removed. Thus, in this case, the filter generates query pairs according to:
 \begin{equation}\small
      \mathcal{W}_{S}(\mathcal{M})= \{ \langle q_x,q_y-S_{stop} \rangle \in \mathcal{M}\}
 \end{equation}

\section{Research Questions}\label{sec:RQ}

\pageenlarge{2} \noindent This paper focuses on addressing \xa{four} research questions on our \xa{generative query rewriting models:}
\xa{Firstly, we investigate the effectiveness of the models under GenQR framework, namely the T5QR and FlanQR models. In particular, since the choice of training data is key to our fine-tuned T5-based reformulation models, we also investigate the best training settings among the weak supervision filters proposed in Section~\ref{weak_supervision}:}

% \noindent \textbf{RQ1:} \xa{How are the performance of the query reformulation models under GenQR framework, namely T5QR and FlanQR models? }

\noindent \textbf{RQ1:} \xa{What is the impact of the weak supervision training techniques for GenQR models? }

% Firstly, \io{since} the choice of training data is key to \xa{our fine-tuned} T5-based reformulation models, we
% % firstly
% investigate the best training settings among the weak supervision filters proposed in Section~\ref{weak_supervision}:

% \noindent \textbf{RQ1:} What is the impact of the weak supervision filters \iadh{when} applied to the initial training pool on \inote{do we really measure training time?} \craigc{T5 fine-tuning time and} the retrieval effectiveness of the \craigc{resulting} \cm{T5QR} generated paraphrases?

Secondly, since we propose \xa{GenPRF}, which makes use of contextualised input, we present our second research question:

\noindent \textbf{RQ2:} \xa{How does the additional pseudo-relevance contextual information affect the performance of GenPRF compared to GenQR, specifically when using T5PRF vs. T5QR and FlanPRF vs. FlanQR?}
% Does additional pseudo-relevance \iadh{contextual} information \xa{brings gains or downsides for GenPRF than GenQR, i.e. T5PRF vs. T5QR and FlanPRF vs. FlanQR?}
% ensure that T5PRF is more effective than T5QR.

Thirdly, we compare the effectiveness of \io{the studied} \xa{generative reformulation models, including both T5-based and \xa{FLAN-based},} with standard and recent query reformulation baselines by addressing the \iadh{following} research question:

\noindent \textbf{RQ3:} How \io{do the studied} \xa{generative} reformulation models perform compared \iadh{to the} baseline query reformulation models?

Next, we investigate the effectiveness of the \xa{generative}
reformulation models within an advanced neural re-ranking pipeline. \iadh{Hence,} we propose the \iadh{following} \io{fourth} research question:

\looseness -1 \noindent \textbf{RQ4:} Do queries reformulated using \xa{generative reformulation} models result in further improvements when combined with a neural re-ranker?
%other enhanced retrieval approaches such as ColBERT?
%\inote{does ColBERT come from nowhere? SHoudl ColBERT not be part of the exp setup}

% \looseness -1 \xa{Finally, we examine the effectiveness between the fine-tuning and prompting methods for GenQR and GenPRF frameworks.}

% \noindent \textbf{RQ5:} \xa{How do the prompting-based models, FlanQR and FlanPRF, perform compare to the fine-tuning based models, T5QR and T5PRF?  }
% \inote{added as discussion section}

% In particular, we have similar parameters to existing query expansion models, such as the weight of the reformulated query compared to the user's initial query (sometimes called Rochhio's $\beta$) during in inverted index retrieval. By mapping this parameter to T5PRF, we arrive at the following research question:
% \inote{move hyperparameter study to appendix}
% Finally, we aim to determine the \io{effect} of \io{the} hyperparameters in our T5-based reformulation models. 
% \noindent \textbf{RQ5:} What is the impact of the number of top $M$ passages in T5PRF, \xiaow{the number of paraphrases $N$ to form the query reformulations \craigc{as well as the relative weighting of reformulations obtained from RM3 and T5-based reformulation models}?}

\section{Experimental Setup}\label{sec:experimental_setup}
\looseness -1 In this section, we \iadh{present} the \iadh{datasets} used in our experiments in Section~\ref{dataset}. Then, we describe the \xa{implementation details for GenQR and GenPRF models in Section~\ref{ssec:T5_setup} and present the configuration of the retrieval pipeline in Section~\ref{ssec:retrieval_setup}.}
% including the fine-tuning of the \xiaow{pretrained }T5 model and the configuration of \iadh{the} retrieval pipeline in \iadh{Sections}~\ref{ssec:T5_setup} \& ~\ref{ssec:retrieval_setup}. 
Finally,  Section~\ref{ssec:baselineQR} \iadh{provides details about the used} \craigc{baselines}.
%pre-process for the training data:

\subsection{Datasets}\label{dataset}
\looseness -1 \io{As discussed before, one} challenge when \craigc{fine-tuning} the T5 model for query reformulation is that there \iadh{are} no gold-standard labelled \craigc{pairs of queries} \craigc{that indicate improved query formulations}.  To circumvent the lack of ground-truth data, we leverage the filters proposed in Section~\ref{weak_supervision} to generate the required weak supervised query pairs to fine-tune the T5 model, instead of \io{using} human annotated query pairs. In our work, the training dataset is constructed based on the MSMARCO document ranking dataset\footnote{https://microsoft.github.io/msmarco/} as used in the \craig{recent} TREC Deep Learning tracks. The MSMARCO training dataset contains $\sim$3.2M documents along with 367K training queries, each with 1-2 labelled relevant documents.

We follow the assumption introduced by Zerveas et al.~\cite{zerveasbrown} that if a document is labelled as relevant for multiple queries, such queries are assumed to convey the same information need. Thus, we firstly identify 188,292 pairs of training queries that share the same labelled relevant document(s), then we further apply the filters introduced in Section~\ref{weak_supervision} to reduce the noisy pairs, and increase the likelihood that they represent \io{near-identical} information needs.

\looseness -1 For the \craigws{Stopwords filter}, we use a stopwords list of 733 words obtained from the Terrier IR platform.
%\craig{of} Terrier \xiao{v5.3}~\cite{ounis2005terrier}. 
\craig{For} the overlap \craigws{filter}, the top-K \iadh{parameter} is set \iadh{to} 10 \xiao{as suggested in~\cite{mass2020unsupervised}} and the threshold \iadh{value} is configured as $\delta_{O}=5$.
% $\delta_{O}=\{1,3,5\}$ ($\delta_{O}=3$ \craig{was suggested by}~\cite{mass2020unsupervised}). Thus, with this setting, we generate three pools with \iadh{the} overlap \craigws{filter} applied. 
For the \craigws{effectiveness filter  $\mathcal{W}_{E}$}, we \craigws{remove} training pairs from \iadh{the} \xiao{initial pool} based on the difference in the discounted cumulative gain (DCG), using a minimum difference thresholds of $\delta_{E}=0$. Statistics of the resulting pools after applying each filter are \craig{shown} in Table~\ref{tab:summary_training}. \craig{We do not select any more aggressive filter settings (cf.\ $\delta_{O} > 3$ or $\delta_{E} > 0$) as this results in pools smaller than 20-30k query pairs, \iadh{which} we found during \iadh{our} initial experiments \iadh{to result in} lowly performing query reformulation models.}
%\inote{check with table 2 and table3 do we include all the WS filters?}}
% and (2) $\delta_{E}=0.1$

%--  for instance, the content of a document might address multiple different topics, or the queries pertain to different information needs. In particular, we apply a filter named as overlap filter, which assumes that only query pairs for which there is a marked overlap in the retrieval documents for both queries are suitable training examples~\cite{cronen2004framework,cronen2004language}. On top of that, as many query pairs may consist of syntactical variations of the same query, such as moved or changed stopwords. Thus, in order to help the pretrained T5QR model focus on generating terms that can identify relevant documents, we apply a filter named as Stopwords Filter, which refines the generated training pairs conditioned on the presence of the common words or stopwords.

\begin{table}[tb]
    \centering
    \caption{Summary of query pair training pools.}
    \vspace{-1\baselineskip} \label{tab:summary_training}  
     \resizebox{85mm}{!}{
    \begin{tabular}{lllrrr}
\toprule 
Method  & Description & Pool Size & AvgLen($q_x$) & AvgLen($q_y$) \\
\midrule
 {Initial pool} &$ Relevance(q_x)=Relevance(q_y)$ &188,292 &6.44  &6.44\\

% \hline
\multirow{1}{*} {$\mathcal{W}_{O}$ } 
% &$\mathcal{W}_{O1}$&  $\delta_{O}=1$
%  & 105,488 &6.41 &6.41\\ 
% &$\mathcal{W}_{O2}$ & $\delta_{O}=3$ & 63,742  &6.27 &6.27 \\ 
 & $\delta_{O}=5$ &40,016   &6.08 &6.08 \\

%\hline 
\multirow{1}{*} {$\mathcal{W}_{E}$ } 
&  $\delta_{E}=0$     & 64,009  &6.52 &6.63\\ 
% & $\mathcal{W}_{E2}$&  $\delta_{E}=0.1$  & 31,733  &6.47 &6.70 \\ 
% & $\mathcal{W}(DCG)_{E3}$& $\delta_{E} > DCG(q_x)\times 10\%$ & 52,789 &6.53 &6.69\\
%\hline 
\multirow{1}{*} {$\mathcal{W}_{S}$ }
 
 &  $\langle q_x,q_y-S_{stop} \rangle $      & 188,292  &6.44  &3.46\\ 
% \todo[xiao]{Q1: what name would be suitable?
\bottomrule
\end{tabular}
}\vspace{-1\baselineskip}
\end{table}
 
% \subsection{Evaluation Dataset}
\pageenlarge{1} \looseness -1 We evaluate the retrieval effectiveness of our proposed \xiao{T5QR} framework on four standard test collections, \craigc{namely: the TREC 2019 Deep Learning track~\cite{craswell2020overview} (i) document ranking \xiaow{ and (ii) passage ranking tasks} (both containing 43 queries), (iii) 250 queries from \io{the} TREC Robust 2004 track~\cite{voorhees2004overview}, \io{and} (iv) \xiaow{149 description-only} queries from the TREC Terabyte Track using GOV2~\cite{clarke2004overview}}. 
% and 100 TREC Web track topics using WT10G~\cite{chiang2005wt10g}. 
\xiaow{For \io{the} Robust 2004 test collection, we experiment using both the title-only and the description-only queries, to allow further comparisons with existing \craigc{approaches from} the literature.} %\inote{unclear how you reached 4 test collections}.}

% For the latter 3 test collections, we experiment using the description-only queries -- as also done by~\cite{li2018nprf}. Indeed, the description is a sentence describing the information need rather than the keyword-only versions found in the title field, as these are closer to the question-type queries found in MSMSARCO\footnote{Later, in Section~\ref{ssec:RQ4}, we also conduct experiments on Robust04 using the title-only topics, to allow further comparisons with existing works in the literature.}.\inote{update this para}

%We employ 250 description queries of Robust04. \iadh{In addition}, we evaluate our model using 150 and 100 description \iadh{queries} from GOV2 and WT10G, respectively. 
% \inote{justify} \inote{also remind what title description queries means}
% ,  we use the test queries from the TREC 2019  Deep Learning tracks, with 43 queries respectively.
% In addition, we \inote{Robust04, GOV2, WT10G}

% three standard test collections, namely test collections from TREC 2019 \& 2020 Deep Learning tracks with 43 \& 45 queries respectively. 

% Table~\ref{tab:summary_training} (first row) reports the salient statistics of the initial training pool.
%

%   
\subsection{Implementing GenQR and GenPRF}\label{ssec:T5_setup}

\textbf{Fine-tuning the T5 Models:}
% For our experiments, as an implementation for the T5 model, 
\xa{When implementing T5QR and T5PRF}, we use the
\xa{{\em t5-base}} model with \xa{220 million} parameters~\cite{raffel2020exploring} \craigc{obtained via} the HuggingFace transformers library\footnote{https://github.com/huggingface/transformers}.
Following~\cite{raffel2020exploring}, fine-tuning is performed using a learning rate of $3 \times 10^{-4}$ and a dropout rate of $10\%$. \craigc{We train for 4 epochs, using a batch size of 6.}

For \io{the} configuration of the T5PRF \io{model} using contextualised inputs, the \craig{size of the pseudo-relevance feedback set, $K$, is set to 10\xiao{, following~\cite{li2018nprf,zheng2020bert}}.}
A sliding window with a size of 128 and a \craigc{stride} of 64 \craigc{tokens} is used to split each feedback document into passages\xiao{, following \cite{suuniversity}.} 
We select $M=\{1,2,3\}$ passages for use \io{by} the FirstP, TopP or MaxP \io{selection approaches}  (cf.\ Equations~\eqref{eq:PRF-FirstP} -- \eqref{eq:PRF-TopP}) as context input for our T5PRF model - this ensures that the maximum input length of the T5 model is not exceeded.  

\looseness -1 To generate the prediction sequence, the greedy decoding method is used. We use beam search with a beam size of \xiaow{100 to form a large enough set of}
% generate 100 
candidate paraphrases for each input query.
% \inote{why 100?}.
Then we rank all the candidate paraphrases according \craigc{to their likelihood}, 
% calculated using Equation~\eqref{equ:weight},
and select the $N$ most likely paraphrases to form the reformulated query. In our experiments, we use $N=5$.
% , but return to the selection of $N$. 
\xa{More details of the fine-tuning process for T5-based models and the hyperparameter-tuning are provided in Appendix~\ref{app:fine_tune_process} and Appendix~\ref{app:hyper_param}, respectively.}
% \inote{TQ5 surely?}.}

% Then, Sentence-BERT\footnote{https://github.com/UKPLab/ sentence-transformers}~\cite{reimers2019sentence} is employed to select the best paraphrase among the 10 generated candidates. In particular, Sentence-BERT measures the semantic similarity between two sentences by comparing the cosine-similarity of their sentence embeddings to avoid topical drift.\footnote{Initial experiments found Sentence-BERT to be better at identifying high quality paraphrases than a random selection, \cm{as it models the coherence of the generated paraphrases}.}
 
 \textbf{\xa{Prompting the FLAN-T5 Models:}}
 % model version xxl
 % detail the prompts used for each version
\xa{When implementing FlanQR and FlanPRF models, we employ the {\em flan-t5-xxl} model with 3B parameters~\cite{wei2021finetuned}.}\footnote{The models' sizes are selected based on our available computational capacity. Since T5QR and T5PRF require training, we were unable to use as large of a model as we were for the inference-only FlanQR and FlanPRF.}
% to ensure the relatively strong reasoning ability exhibited by larger language models~\cite{weiemergent, huang2022towards}.}
% \footnote{We choose the xxl size to ensure a relatively good reasoning ability exhibit in larger language models~\cite{}.}}
% ``flan-t5-xxl'' 
The configuration for extracting the pseudo-relevance feedback context information for FlanPRF is same as the T5PRF. 
For an input query, we design task-aware prompts for FlanQR as follows,
\texttt{FlanQR}(`\texttt{Improve the search effectiveness by suggesting expansion terms for the query:} {input query}'). The prompts for FlanPRF conditioned on the input query as well as its contextual information, $\texttt{context(R)}$ is as follows,
\texttt{FlanPRF}(`\texttt{Improve the search effectiveness by suggesting expansion terms for the query:} {input query}, \texttt{based on the given context information:} {context(R)}'). \xa{The selected prompts were identified among 20 candidate prompts based on their validation performance on the MSMARCO TREC 2019 passage ranking query set. }
% \inote{motivation}}
% \inote{I would indeed like more of a story about how we arrived at these prompts}
% \inote{risk of giving too specific prompts? what kind of information we would expect FLan to produce using a particular word, e.g. we include the "improve search effectiveness" to specify the usage of these expansion terms. }

% effectiveness of T5PRF, which integrates contextual input in the form of important passages from a pseudo-relevant feedback set. 
% \todo{SM: can we somehow motivate the discrepancy between model sizes?}

\subsection{Retrieval Pipeline Setting}\label{ssec:retrieval_setup}

%We conduct retrieval experiments on the PyTerrier\footnote{https://github.com/terrier-org/pyterrier} platform~\cite{macdonald2020declarative} using its ColBERT integration\footnote{https://github.com/cmacdonald/pyterrier$\_$bert}. The index is built by Terrier~\cite{ounis2005terrier}, with stopwords removed but no stemming applied.

\looseness -1 For ranking, we use a Porter stemmed index with stopwords removed.  We \craigc{apply} a two-stage ranking pipeline, where the documents are first ranked by \xiaow{a tuned BM25 retrieval model \xiaow{using grid search.}
% \inote{how it is tuned?}. 
In the second stage, \io{re-ranking} is \craigc{performed} by 
% vanilla BERT~\cite{macavaney2019cedr,nogueira2019passage} and
\xiaoJ{the monoT5~\cite{pradeep2021expando} model}.} 

%\inote{citation}}
% the \iadh{parameter-free} DPH retrieval model\footnote{\cm{DPH performs similarly to BM25 without need resort to and time-consuming tuning of parameters in the first-stage of a ranking pipeline - later, in Table~\ref{tab:RQ4}, we also provide BM25 results for comparison purposes.}} and then re-ranking, \iadh{in the second stage}, is done by the ColBERT model~\cite{khattab2020colbert}. 
To deploy \xiaoJ{the monoT5 neural re-ranker}, a sliding window is used to break up long documents into passages of 128 tokens with a stride of 64 between passages. \iadh{We} use MaxPassage~\cite{dai2019deeper,liu2002passage} to \craigc{obtain} the final score of a document. %\inote{As mentioned in Section 5.2?} \xiaow{We select 5 generated paraphrases for each given input query to form the reformulated query. }

\def\mix{\beta}
\pageenlarge{2} 
\xa{For the T5QR and T5PRF generated queries,} \xiaow{we combine the generated query reformulation \io{with} the original query to form a new query for input in the first-stage retrieval, i.e.\ $q^{\dag}= k_{RM3}\cdot q^r_{RM3}+ k_{T5}\cdot q^r$, where $k_{RM3} $ and $k_{T5} $ are parameters that control the importance of  reformulations query generated by RM3 or \io{by the} T5-based query reformulation model \craigc{(T5QR or T5PRF)}. We use $k_{RM3}=1$ and $k_{T5}=0.5$ as a default setting, but investigate the impact of these two parameters in addressing RQ5.}
% \inote{surely RQ5?}. }
% a mixing parameter that can control the weight of the reformulated query generated by T5QR or T5PRF model. We use $\mix=0.5$ as a default setting, but investigate the impact of this parameter in addressing RQ3.}
% Finally, we will release our fine-tuned models and code on acceptance of this paper.
% \inote{the prompting retrieval pipeline. we have beta as the weight}
\xa{For the FlanQR and FlanPRF generated queries, we append the generated query reformulations to the original query.\footnote{We tested versions for expansion terms both with and without RM3 terms for both T5 and FlanT5. T5 was more effective \smmm{on our validation data} with interpolated RM3 terms, while FlanT5 was more effective without. Therefore, we report these results.} In particular, the importance of the generated query reformulations \xa{compared to the original query} is controlled by a parameter $\beta$. We use $\beta=0.2$ as the default setting in this work. All the hyperparameters are selected based on the validation performance on the TREC 2019 passage ranking query set.}

%\inote{anonymisation}   
%\looseness -1  We implement a two-stage ranking pipeline, where the documents are first ranking by the DPH retrieval model and then re-ranking is done by the ColBERT model~\cite{khattab2020colbert} in the second stage. To deploy ColBERT for document re-ranking, a sliding window is used to break up a long document into passages of 128 tokens with a stride of 64 between passages and use MaxPassage~\cite{dai2019deeper,liu2002passage} to give the final score of a document. 

% \looseness -1 Inspired by typical query expansion models from Rocchio onwards, we combine the generated query reformulation to the original query to form a new query for input in the first-stage retrieval, i.e.\ $q^{\dag}=q^0+\mix\cdot q{\prime}$, where $\mix $ is a mixing parameter that can control the weight of the reformulated query. We use $\mix=1$ as a default setting, but investigate the impact of this parameter in addressing RQ3.
% Similarly, during the second stage of the retrieval pipeline, which is used when addressing RQ3, we score \iadh{the} original and reformulated queries separately using ColBERT, using a parameter $\varphi$ to control \iadh{the} emphasis on the reformulated query, i.e.\ $re-ranker(q,d) = ColBERT(q^0,d) +\varphi ColBERT(q\prime, d)$. 

% Finally, we will release our fine-tuned models and code to use them on acceptance of this paper.

\subsection{Baseline Models}\label{ssec:baselineQR}
In order to \iadh{evaluate} the effectiveness of our proposed \xa{generative query reformulation models,}
% T5QR and T5PRF models in generating reformulated queries,
we compare \io{them to} \xiaoJ{six families of query reformulation baselines, namely:}
% various reformulation baselines, namely:

% \begin{itemize}
 {\em Initial query:} The original query without any modification.

\looseness -1 {\em Traditional Query Expansion.}
Three traditional query expansion models are used as baselines: {(i) BM25+RM3~\cite{abdul2004umass}}, {(ii) BM25+Bo1~\cite{amati2003probability}} and {(iii) BM25+KL~\cite{amati2003probability}}. We use BM25+RM3 as our main PRF baseline, due to its suitability as a baseline for neural methods~\cite{lin2019neural}.

\looseness -1 {\em Neural Query Reformulation.}
{(i)} The Transformer model was proposed by Vaswani et al.~\cite{vaswani2017attention}, and was employed by Zerveas et al.~\cite{zerveasbrown} for the query reformulation task. We implement the transformer model using the OpenNMT platform~\cite{klein2017opennmt}; {(ii) Sequence-to-Sequence Model with Attention.} This model consists of an RNN-based encoder and decoder with an attention mechanism. The model is also implemented using the OpenNMT platform~\cite{klein2017opennmt}; {(iii) } GPT2~\cite{radford2019language} is a pretrained transformer-based language model. We fine-tune the GPT2 model for \iadh{the} query reformulation task as a baseline model.

\begin{table*}[tb]
\caption {\looseness -1 \xa{RQ}1: Comparison between weak supervision filters. Superscripts a/b denote significant improvements (paired t-test with Holm-Bonferroni correction, $p < 0.05$) over the indicated baseline model. The highest value in each column is boldfaced. 
% \inote{in terms of recall, we choose E+S as the filter for the following research questions}
}\vspace{-1\baselineskip} 
\centering  
\def\ttestB{$\dagger$}
\resizebox{180mm}{!}{
\begin{tabular}{ll cccc cccc ccc ccc ccc}

%  \hline \hline 
 \toprule
 \multirow{2}{*}{ \bf{QR Models}}  & \multirow{2}{*}{\bf{Training/epoch}} &  \multicolumn{4}{c }{\bf{TREC 2019 Document}} & \multicolumn{4}{c}{\bf{TREC 2019 Passage}} &\multicolumn{3}{c}{\bf{Robust04 (T)}} &\multicolumn{3}{c}{\bf{Robust04 (D)}}& \multicolumn{3}{c}{\bf{GOV2}}   \\
\cmidrule{3-19}

   & &  MAP  & MRR & R@1k & nDCG@10&  MAP  & MRR & R@1k & nDCG@10   &  MAP   & R@1k & nDCG@10  &  MAP   & R@1k & nDCG@10 &MAP  & R@1k & nDCG@10\\
  
  \midrule
  Initial query (a) &-
  & .341 & .889 & .701 &.557
  & .311 & .694 & .752 &.517
  
  & .256  &.705 &.439
  & .233 &.670  &.411
  & .268  &.643 &.468
  \\

  BM25+RM3 (b)&-
  & .397 & .858 &.760 & .559
  & .342 & .655 &.796 & .548
  & .284  &.752 & .437
  & .266  &.709 & .427
  & .291  &.657 & .452
  \\ 
  
  \midrule

%   KPtwo-IDFBatch 
 $Initial Pool$ &  168 min
  & .397$^{a}$ & .849 & .757 & .565
  & .347$^{a}$ & .663 &.796$^{a}$  &.554
  & .286$^{a}$   &.754 &.443
  & .271$^{ab}$  & .721$^{ab}$ & .434$^{b}$
  & .293$^{ab}$  & .662$^{b}$ &.461
 
  \\
                           
  $W_S$  &  163 min %case2 
  & .387$^{a}$ & .833 & .745 &.579
  & .342 &.654 & .777 &.532
  & \bf{.290}$^{ab}$  &\bf{.760} &.444
  & \bf{.279}$^{ab}$  &.736$^{ab}$ &.439
  & .301$^{ab}$  &.680$^{ab}$ &.474
  \\

  $\mathcal{W}_{O}$ & 35 min% case73
  &.398$^{a}$ &.860 &.756$^{a}$  &.581
  &.343 &.644 &.797$^{a}$ &.543
  &.285$^{a}$  &.753 &.442
  &.268$^{ab}$  &.711$^{ab}$ &.428
  &.294$^{ab}$  & {.663}$^{b}$ &.463
  \\

  $\mathcal{W}_{E}$ & 55 min %case 62
  &\bf{.400}$^{a}$ &.856 &.761$^{a}$ & .581 
  &.348$^{a}$ &.657 &.808$^{a}$ &.547
  &.286$^{a}$ &753 &.442
  &.269$^{ab}$  &.714$^{ab}$ &.429
  &.292$^{ab}$  &.658$^{b}$ &.460
  \\
\midrule
$\mathcal{W}_{(O+S)}$ & 34 min 
% $W_{(O3+S)}$ case 732
&.397$^{a}$ &.875 & .756$^{a}$ & .580
&.344       &.634  &.793 &.542
&.286$^{ab}$  &.744 &. 442
&.270$^{ab}$  &.718$^{ab}$ &.429
&.302$^{ab}$  &.675$^{ab}$ &.472$^{b}$
\\

$\mathcal{W}_{(E+S)}$ &  54 min
% \spadesuit case 622
&.398$^{a}$ & \bf{.914} &.753$^{a}$ & \bf{.600 }
&.351$^{a}$ &.645$^{a}$ &.788 &.535
&.287$^{ab}$  &.753 &.442
&.278$^{ab}$  &.734$^{ab}$ &.439%$^{ab}$
&.302$^{ab}$  &.676$^{ab}$ &.474$^{b}$
\\
\midrule
$FlanQR$ &- 
&.384$^{a}$ &.790 &\bf{.763}$^{a}$ &.537
% &\bf{.521} &\bf{.908} &\bf{.845} &\bf{.727}
&\bf{.382}$^{a}$ &\bf{.707} &\bf{.845}$^{a}$  &\bf{.556}
&.270$^{ab}$ &.756$^{a}$ &\bf{.461}
&.278$^{a}$ &\bf{.760}$^{ab}$ &\bf{.471}$^{ab}$
&.\bf{303}$^{ab}$ &\bf{.680}$^{ab}$ &\bf{.510}$^{ab}$

\\
% \midrule
% $FlanQR-merged$ &
% &&&&
% &.349 &.667 &.823 &.516
% \\

\bottomrule
\end{tabular}
}%\vspace{}%-\baselineskip}
% }
% \end{ruledtabular}
\label{tab:RQ1}\vspace{-1\baselineskip}
\end{table*}
%%%%%%%%%%%%%%%%%%%%%%

\looseness -1 {\em Neural Document Expansion.}
\noindent{ DocT5query\xiaow{~\cite{nogueira2019doc2query} is a document expansion model, which uses a T5 model fine-tuned \craigc{to predict related queries for a given document.} These are then appended to the original document during indexing time.}} \xf{We obtain the results of DocT5query for the MSMARCO document corpus from~\cite{ma2022document}.}
% \xf{We compare the results of DocT5query for the MSMARCO document corpus reported in this work~\cite{ma2022document}.}
% \craigc{We index the DocT5query predicted queries for the MSMARCO document corpus produced by the original authors; we note that no DocT5query predicted queries are known to exist for Robust04 nor \io{nor for the larger} GOV2 corpus, possibly due to the expense in analysing the latter large corpus.}  

\looseness -1  {\em \xiaoJ{CEQE Models.}}
% makes use of BERT \xiaow{(Large)} for language representation, as well as \inote{cannot parse this sentence} several neural re-ranking stages (e.g.\ neural re-ranking is applied for selecting pseudo-relevance feedback passages).
\xiaoJ{CEQE is a BERT-embedding based query expansion model. In particular, 70-100 expansion terms selected in the contextualised embedding space are added to the original query to form a new query.  We \craigj{compare} with three variants of CEQE, namely CEQE-Max, CEQE-Centroid and CEQE-mul models.  In our implementation, we apply CEQE query expansion models upon the documents retrieved by BM25 and apply the pipeline BM25 + RM3 + BM25 rather than the Dirichlet LM + RM3 +BM25 pipeline which is used by the original CEQE implementation~\cite{naseri2021ceqe}.}

\pageenlarge{1}{\em Neural Query Expansion Models.}
\noindent{(i) NPRF}~\cite{li2018nprf} is a neural pseudo-relevance feedback model, which operates as a re-ranker, that it does not generate a refined textual query that is re-applied to the inverted index, but instead uses the pseudo-relevance feedback to re-ranked the initial candidate set of documents. The best variant -- NPRF$_{ds}$-DRMM -- is included as a baseline; (ii) BERT-QE~\cite{zheng2020bert} is also a neural PRF model that builds on NPRF, but use BERT \xiaow{(Large)} to refine the PRF information. \xiaow{We compare with the reported best variant, BERT-QE-LLL.} 

% \inote{add the query2doc???}

% \craigc{Its best variant is CEQE-MaxPool, which expands the query with upto 100 expansion terms}.}

%\inote{we compare with which variant of BERTQE}
% \end{itemize}
% For a fair comparison, the 

% \todo[iadh]{The latex could be beautified}

% \subsection{Measures}

% \bf Evaluation Metrics:
\pageenlarge{1}
\subsection{Evaluation Metrics}\label{evaluation_metrics}
% \pageenlarge{2} 
Following standard practice \iadh{in} the TREC Deep Learning track~\cite{craswell2020overview}, we measure the effectiveness of the reformulated queries through their ranking performance in terms of Mean Average Precision (MAP) and Mean Reciprocal Rank (MRR)\footnote{Although there has recently been some discussion about the choice of the MAP and MRR metrics to report in the IR community, no consensus has yet been reached. Thus, to compare with the existing works, we follow the widely used metrics in our work.}, as well as \xiaow{Recall and} normalised discounted cumulative gain calculated to rank depth 10. 
% We use the same measures on \iadh{the} GOV2, Robust04 and WT10G test collections. 
%, 20 \& 1000.
% \inote{Metrics used to compare with existing baselines}
We use the paired t-test ($p<0.05$) for significance testing and apply \iadh{the} Holm-Bonferroni multiple testing correction.

 \section{Results and Discussions}\label{sec:results}

\looseness -1 \craigc{We now} address our \xa{four} research questions in turn: the impact of the weak supervision filters on training data quality \xa{for GenQR models} (Section~\ref{ssec:RQ1}); the usefulness of pseudo-relevance feedback as contextualised input, as per our proposed \xa{GenPRF} model (Section~\ref{ssec:RQ2}). \xiaow{Then, we make comparisons with \io{the} baselines (Section~\ref{ssec:RQ3})} and consider the role of a neural re-ranker (Section~\ref{ssec:RQ4}). 
% Finally, we investigate the role of the hyperparameters in attaining effective retrieval (Section~\ref{ssec:RQ5}).
% Finally, we consider the role of a neural re-ranker (Section~\ref{ssec:RQ4}) and make comparisons with baselines (Section~\ref{ssec:RQ3}).

\subsection{RQ1: Impact of the Training Data Quality}\label{ssec:RQ1}
% \inote{revisit}

\begin{table*}[tb]
\caption {RQ2: Effect of \io{the} PRF contextualised input. Superscripts a/b/c/d  denote significant improvements (paired t-test with Holm-Bonferroni correction, $p < 0.05$) over the indicated baseline model(s). The highest value in each column is boldfaced. }\vspace{-1\baselineskip}  
% \inote{!!!robust TD, GOV2, FlanPRF all lower than bm25}}
% \inote{!!: only can improve on trec psg, check trec document MRR, sth wrong there: conclusion: topp and fistp better than maxp, and we take the topp in the following sections}
% }   
\centering
\def\ttestB{$\dagger$}
\resizebox{180mm}{!}{
\begin{tabular}{l cccc cccc ccc ccc ccc}
%  \hline \hline 
 \toprule
 \multirow{2}{*}{\textbf{QR Models}}
 &\multicolumn{4}{c}{\textbf{TREC 2019 Document}} 
 &\multicolumn{4}{c}{\textbf{TREC 2019 Passage} }
 &\multicolumn{3}{c}{\textbf{Robust04 (T)} }
 &\multicolumn{3}{c}{\textbf{Robust04 (D)}}
 &\multicolumn{3}{c}{\textbf{GOV2} }  \\
\cmidrule{2-18}

  % {QR Models}  
  &  MAP   & MRR & R@1k & nDCG@10
  &  MAP   & MRR & R@1k & nDCG@10   
  &  MAP   & R@1k       & nDCG@10  
  &  MAP   & R@1k       & nDCG@10 
  &  MAP   & R@1k       & nDCG@10\\
  
  \midrule
  Initial query (a)
 & .341 & .889 & .701 &.557
& .311 & .694 & .752 &.517
& .256  & .705 &.439
&.233 &.670 &.411
& .268 &.643 &.468
  \\

  BM25+RM3 (b) 
  &.397   &.858  &.760 &.579
  &.342   &.655  &.796 &.548
  &.284   &.752  &.437
  &.266   &.709  &.427
  & .291  &.656  &.452
  \\ 
  
  % \midrule

\midrule
$T5QR$ (d)
&.398$^{a}$   &\bf{.914}   &.753    &.600
&.351         &.645        &.788    &.535
&.288$^{a}$   &.756$^{a}$  &.443
&.277$^{ab}$  &.734$^{ab}$ &.439$^{ab}$
&.302$^{ab}$  &.676$^{ab}$ &.474$^{b}$

\\
$T5PRF_{FirstP}$ 
&\bf{.411}$^{a}$  &.868          &.770     &.596
& - &-&-&-
&\bf{.297}$^{ab}$ &.757$^{a}$    &\bf{.463}$^{bd}$ 
&.282$^{abd}$     &.740$^{abd}$  &.441$^{abd}$
&.302$^{ab}$      &.676$^{ab}$   &.475
\\
$T5PRF_{TopP}$     %\spadesuit
&.411$^{a}$       &.868              &{.776}$^{a}$   &{.602}$^{a}$
&.353$^{a}$  &{.676}            &.815$^{a}$     &.554%topPTwo622 for psg
&.296$^{ab}$      &\bf{.759}$^{a}$   &.461$^{bd}$ 
&\bf{.285}$^{abd}$     &.743$^{abd}$      &.447$^{abd}$
&\bf{.303}$^{ab}$ &\bf{.685}$^{ab}$       &.479$^{b}$

\\

$T5PRF_{MaxP}$
&.411$^{a}$       &.868             &{.776}$^{a}$    &{.602}$^{a}$
& - &-&-&-
&.296$^{ab}$      &{.759}$^{a}$  &.461$^{bd}$ 
&\bf{.285}$^{abd}$     &.743$^{abd}$     &.447$^{abd}$
&\bf{.303}$^{ab}$ &\bf{.685}$^{ab}$      &.479$^{b}$
\\

\midrule
$FlanQR$ (e)  
&.384$^{a}$ &.790 &.763$^{a}$ &.537
% &\bf{.521} &\bf{.908} &.845 &\bf{.727}
&.382$^{a}$ &.707 &.845$^{a}$  &.556
&.270$^{ab}$  &.756$^{ab}$  &.461$^{ab}$ 
&.278$^{a}$  &\bf{.760}$^{ab}$  &\bf{.471}$^{ab}$
&\bf{.303}$^{ab}$ &.680$^{ab}$  &\bf{.510}$^{ab}$ 

\\
$FlanPRF_{FirstP}$ 
&.373 &.816 &.841 &\bf{.605}
&- &- &-&-
&.260  &.699  &.434
&.225  &.656  &.433$^{b}$ 
% &.262 &.598 &\bf{.485}
&.258 &.597 &.490$^{b}$
\\ 
$FlanPRF_{TopP}$ 
&.373 &.816 &.841 &\bf{.605}
&\bf{.404}$^{ab}$ &\bf{.809}$^{b}$ &\bf{.866}$^{ab}$ &\bf{.628}$^{ab}$
&.260 &.699 &.434
&.224 &.656 &.406 
&.258 &.597 &.490$^{b}$
\\
$FlanPRF_{MaxP}$ 
&.361 &.780 &\bf{.844} &.578
&- &- &-&-
&.243 &.683 &.420
&.225 &.656 &.433$^{b}$ 
% &.261 &.583 &.471
&.257 &.597 &.496$^{b}$
\\
\bottomrule
\end{tabular}}%  

% \end{ruledtabular}
\label{tab:RQ2}\vspace{-1\baselineskip}
\end{table*}
%%%%%%%%%%%%%%%%%%%%%%%

In order to better fine-tune the T5QR model, in Section~\ref{weak_supervision}, we proposed to generate refined training pools using different weak supervision filters. Table~\ref{tab:RQ1} compares the \xiaow{average training time needed per-epoch and the} \craigc{effectiveness} of T5QR trained with three different weak supervision filters, namely the Stopwords Filter, \io{the} Overlap Filter and \io{the} Effectiveness Filter (top half), as well as \io{using} combinations of these filters (bottom half). \xa{In addition, we also report the retrieval effectiveness for FlanQR model in Table~\ref{tab:RQ1}, which does not require fine-tuning.} We report results on all four test collections: TREC 2019 \xiaow{for document ranking and passage ranking tasks, \craigc{as well as} GOV2 and Robust04}.

\pageenlarge{2} From Table~\ref{tab:RQ1}, we \craigc{observe} that both T5QR trained with \io{the} initial pool and the smaller pools result in significant improvements over the initial query on all datasets and significantly \io{outperforms} BM25+RM3 for a few  \craig{settings and measures, particularly on Robust04 (D) and GOV2}. In addition, we also note that applying \io{the} Overlap filter and \io{the} Effectiveness filter only \io{takes} \craigc{33\% and 20\%} of the training time needed for the initial pool, respectively.
Among the filters, the performances achieved are similar, and hence we conclude that there is no need to use the initial pool as the training time is less with more aggressive filters. \xa{Moreover,}
% In addition,
we observe that when combining with the stopwords filter, these combined filters demonstrate marked performance enhancement over the single filters. \xa{Thus, to avoid spending time training the T5 model to generate useless stopwords that do not affect the BM25 retrieval process, 
% \xa{more importantly, in the retrieve-then-rerank pipeline, T5QR is expected to exhibit higher recall. Thus, 
we take forward the higher recall combined filter, i.e. $\mathcal{W}_{(E+S)}$, when addressing RQ2.}

\xa{Furthermore, we also report the performance of FlanQR, which does not \xa{require fine-tuning.}
% rely on a training process. 
We observe that FlanQR leads to significant improvements over the initial query, which indicates the usefulness of the prompting-generated query reformulations.}

Overall, in response to RQ1, we find that \xa{both the T5QR and FlanQR} query reformulation models can significantly improve over \io{the} initial query \xiaow{and, in some cases, BM25+RM3.} \xa{For T5QR, using filters can lead to faster training without loss of effectiveness.}
% \sm{while using filters can} lead to faster training \craigc{without loss of effectiveness}. 
 
% \inote{also FlanQR, where we don't need the training data. aim: compare FlanQR with bm25 and rm3}

\begin{table}[tb]
\caption{\looseness -1 RQ3: Comparison with the baseline query expansion approaches.
%as well as with neural query reformulation models using the large initial training pool.
Superscripts a...j denote significant improvements over the indicated baseline model(s).}\vspace{-1\baselineskip}    
% \inote{drop all the psg types no need}
\resizebox{90mm}{!}{
\begin{tabular}{l ccc cc cc cc }
 
 \toprule
 \multirow{2}{*}{\bf{Models}}
 & \multicolumn{3}{c }{\bf{TREC 2019 Doc.}}
 & \multicolumn{2}{c}{\bf{Robust04 (T)}} 
 & \multicolumn{2}{c}{\bf{Robust04 (D)}} 
 & \multicolumn{2}{c}{\bf{GOV2}}\\
  % & \multicolumn{1}{c}{\bf{TREC 2019 Psg.}}\\
 \cmidrule{2-10}
 &MAP 
 &nDCG@10 
 &nDCG@20
 &MAP &nDCG@20 
 &MAP &nDCG@20 
 &MAP &nDCG@20
  % &
 \\

   \midrule

    BM25+Bo1 (a)        
    &.384  
    &\xf{.567}
    &.538  
    &.287   &.433 
    &.271   &.418
    &.295   &.455
    \\ 
    BM25+KL (b)
    &.391  
    &\xf{.570}
    &.558 
    &.290  &.435
    &.271  &.419 
    &.291  &.458
    \\  
    
    BM25+RM3 (c) 
    &.397 
    &\xf{.579}
    &.558  
    &.284 & .427    
    &.266 & .408
    &.291 &..452
    \\
   
    \midrule
  Seq2seq$_{attention}$ (d)
  &.255  
  &\xf{.414}
  & .403
  &.199     & .343 
  &.204     & .343
  &.216     & .377
                          \\  
  Transformer~\cite{zerveasbrown} (e)
 &.238 
 &\xf{.421}
 &.423
 &.214     &.366 
 &.207     &.347
 &.216     &.377
                  	  \\

    GPT2-QR (f)     
  &.327 
  &\xf{.518}
  &.490
  &.237  &.409  
  &.237  &.396
  &.265  &.407
                  	    \\ 
  \midrule
    DocT5query~\cite{nogueira2019doc2query} (g)
     % & .372 &.573
     &-
     % .270     
     &\xf{.597} %https://cs.uwaterloo.ca/~jimmylin/publications/Ma_etal_SIGIR2022.pdf
     &-
     & - & - 
     & - & -
     & - & -
     \\
    \midrule

   CEQE-Max~\cite{naseri2021ceqe}~(h) % the best one)
    %  &\bf{.4161} & .5614 
    %  &.3086 & .4587
    &\bf{.419} 
    &\xf{.554}
    &.541
    &.300 &.425
    &\bf{.290} &.433
    &.283 &.455
     
     \\
     
     CEQE-Centroid~(i)
    %  &0.4144 & 0.5580
    %  &0.3019 &0.4462
    &.417 
    &\xf{.550}
    &.541
    & .299 &.423
    & .289 &.431
    & .286 &.453

     \\
      CEQE-Mul~(j)
    %  &.3724 & .5563
    %  &.2845 &.4360
    &.410 
    &\xf{.535}
    & .541
    &.292 & .415
    &.285 & .424
    &.289 & .448
    
     \\
     
     \midrule
    
        NPRF~\cite{li2018nprf}        
  &-  &-  &-
  &.290     &.450
  &.280     &\bf{.456}
  & - & -
   \\
     
     BERT-QE (LLL)~\cite{zheng2020bert}
     &-   &-  &-
     &\bf{.386} &\bf{.553}
     &-    &- 
     &.268 &\bf{.604}\\ 
    
    % BERT-QE (LLL)~\cite{zheng2020bert}
    %  &-      &- 
    %  &\bf{.3865} &\bf{.5533}
    %  &-    &- 
    %  &.2681 &.6037\\ 
    
  \midrule

$T5QR$ %622
& .398$^{def}$   
&\xf{.600$^{def}$}
& .583$^{def}$
& .288$^{cdef}$  & .432$^{cdef}$ 
& .278$^{bcdef}$ & .420$^{cef}$
& .302$^{cdef}$  & .460$^{cde}$   

\\
   % $T5PRF_{FirstP}$ %622
   % &.411$^{def}$    &.574$^{def}$
   % &.297$^{cdef}$   & .463$^{cdef}$ 
   % &.282$^{abcdef}$ &.423$^{cef}$
   % &.302$^{cdef}$   &.464$^{cde}$
   
   % \\
   $T5PRF$ %$T5PRF_{TopP}$%622
   &.411$^{def}$  
   &\xf{\bf{.605}$^{def}$}
   &\bf{.583}$^{def}$
   &.296$^{cdef}$  &.449$^{cdef}$ 
   &.285$^{abcdef}$ &.428$^{cdef}$
   &\bf{.303}$^{cdef}$       &.465$^{de}$

   \\
   % $T5PRF_{MaxP}$%622
   % &.411$^{def}$  &\bf{.583}$^{def}$
   % &.296$^{cdef}$  &.449$^{cdef}$ 
   % & \bf{.285}$^{abcdef}$ &.428$^{cdef}$
   %  &.303$^{cdef}$        &.465$^{de}$ \\
   $FlanQR$ 
   &.384$^{de}$ 
   &\xf{.537$^{def}$}
   &.520$^{def}$
   &.270$^{def}$ &.436$^{def}$
   &.278$^{def}$ &.446$^{def}$
   &\bf{.303}$^{cdef}$ &.496$^{def}$
   
   \\
   % $FlanPRF_{FirstP}$ 
   % &.373 &.577
   % &.260 &.393
   % &.225 &.656
   % &.262 &.464
   % \\
   $FlanPRF$ %$FlanPRF_{TopP}$ 
   &.373$^{de}$ 
   &\xf{\bf{.605}$^{def}$}
   &.577$^{def}$ 
   &.260$^{def}$ &.411$^{def}$
   &.224$^{def}$ &.393$^{def}$
   &.262$^{de}$ &.464$^{de}$ 
   \\
   % $FlanPRF_{MaxP}$ 
   % &.361 &.549
   % &.243 &.390
   % &.224 &.377
   % &.261 &.453
   % \\
 
 \bottomrule
 \end{tabular}
}  
\label{tab:RQ3}\vspace{-1\baselineskip}
\end{table}
%%%%%%%%%%%%%%%%%%%%

\subsection{RQ2: Effect of PRF Contextualised Input}\label{ssec:RQ2}
% \inote{doing this}
\looseness -1 We now investigate the \xa{effect of the additional PRF contextual information for T5PRF and FlanPRF models.}
% effectiveness of T5PRF, which integrates contextual input in the form of important passages from a pseudo-relevant feedback set. 
In particular, we investigate three selection mechanisms introduced in Section~\ref{context_inputs}, namely \xiao{MaxP, FirstP and TopP}. For the number of context passages we use \xiaow{$M=1$} -- we return to this choice in \xa{Appendix~\ref{app:hyper_param}}. 
% % \xiaow{We also use the filter combinations identified in Section~\ref{ssec:RQ1}}.
% Note that for the TREC 2019 passage ranking task, there is only a single passage to select, which we report as TopP. 
Table~\ref{tab:RQ2} reports the results of our experiments. 
% \craig{Rows} in the table are grouped by training filter setting. 
We again evaluate using all four test collections.

\looseness -1 \xa{First, we analyse the performance of T5PRF models. In Table~\ref{tab:RQ2},} we find that in each group of models, the T5PRF models exhibit some marked improvements over T5QR \xiaow{for all metrics on all query sets, indicating the effectiveness of the contextualised input.}
\xiaow{When comparing to the BM25+RM3 model, marked improvements in terms of MAP, Recall and nDCG@10 are also observed. In particular, \io{the} T5PRF models significantly outperform the BM25+RM3 model in terms of MAP and nDCG@10 on the Robust04 (T),  Robust04 (D), and GOV2 query sets, as well as in terms of Recall on Robust (D) and GOV2. This observation indicates that query reformulations generated by the T5PRF models are capable of retrieving relevant documents that \io{are} not identified by \io{the} RM3 reformulated queries.} 
% \inote{double check!!! examples in appendix}\craigc{This is also apparent in the \io{example} shown in Table~\ref{tab:context_example} - T5PRF focuses on expansion terms very closely related to the definition of \textit{visceral}, such as `viscera', `body'. \io{On the other hand,} RM3 identifies terms more widely related to the body, such as `cardiac', `fat', but which are less likely to identify more relevant documents for this query.}

% \inote{also FlanPRF, where we don't need the training data. aim: compare FlanQR with bm25 and rm3, and compare different types of pages for FlanPRF.}

\pageenlarge{2} 
\xa{Next we analyse the contextual prompting method, FlanPRF. We observe that FlanPRF exhibits higher performance over FlanQR on all the reported metrics for TREC 2019 passage and document query sets, except on MAP for document queries. This indicates the superiority of the additional contextual information for generating more useful expansion terms on these queries. However, on the Robust title \& description queries as well as the GOV2 queries, FlanPRF gives lower performance compared to FlanQR. 
% \inote{does Flan trained on MSMARCO dataset but not on Robust and GOV2 datasets?} 
We postulate that the \xa{FLAN-based} models need more carefully crafted domain-related prompts, perhaps as some of the instruction data for fine-tuning the FlanT5 model comes from the MSMARCO Q\&A training datasets~\cite{wei2021finetuned}.}

\xa{Finally, among the various contextual information selection mechanisms, we find that TopP and FirstP exhibit higher performance than the MaxP method. Additionally, no differences between the TopP and FirstP methods are observed for the T5PRF and FlanPRF models. Therefore, we take forward the TopP method\xa{, where the top-scored chunks among all the feedback information is selected as the contextual input,} for addressing RQ3 and RQ4.}

%\inote{it will be good to show an example of doc retrieved by T5PRF and not by RM3 for a given query; adding a bit of qualitative analysis helps make the section more insigntful than just describing tables}
% using the \xiao{TopP} selection mechanism exhibits an 
% 8\% (0.5572$\rightarrow$0.6018), 13\% (0.5167$\rightarrow$0.5543), 11\% (0.4301$\rightarrow$0.4775) and 12.6\% (0.3583$\rightarrow$0.4035) improvements in nDCG@10 over the initial query on the four datasets in respectively.\xiao{However, for MAP, MRR and NDCG metrics, the absolute best performance in each group are observed in T5PRF models using different context passage selection mechanisms.}
% \inote{to complete}

\looseness -1 Overall, for RQ2, we observe that PRF information, in the form of contextual passages, can \io{bring} further improvements over the plain T5QR model \xa{on all the five test query sets.} \xa{However, T5PRF only benefits from the additional pseudo-relevant information for MSMARCO document and passage query sets and damages the retrieval effectiveness on Robust and GOV2 queries.} 
\xa{Finally, the performance of both T5PRF and GenPRF varies according to the context selection mechanism.}
% \inote{say more about TopP - what does this tell us about context? say it here, or in the previous paragraph}
% However, 
% the performance of T5PRF varies according to the used context selection mechanism and the training filter.
% We further analyse the effectiveness of the context selection mechanism, but focus on using only the  \xiao{$\mathcal{W}_{(E+S)}$} filter \io{when addressing the} remainder of the research questions.% 

% take forward the $T5PRF$ models with $W_{(E+S)}$ filter with the three context selection mechanisms to RQ3.

% \pageenlarge{1}
\subsection{RQ3: Comparison with Baselines}\label{ssec:RQ3}
% \inote{rewrite}
\pageenlarge{1} In this section, \xa{we examine the effectiveness of the GenQR and GenPRF models}, in comparison to the baselines listed in Section~\ref{ssec:baselineQR}, including traditional query expansion models, neural query reformulation baselines, neural document expansion baselines, and neural query expansion models from the literature.
\sm{Due to space constraints,} Table~\ref{tab:RQ3} compares the performances in terms of MAP and nDCG@20 on the three test collections where the Neural PRF and BERT-QE baselines have been evaluated in the literature: TREC 2019, \xiaoJ{GOV2}, \io{as well as the} \io{title-only} (T) and description-only (D) queries of Robust04.\footnote{For Robust04, we report nDCG@20 to allow comparisons to be made to~\cite{li2018nprf,zheng2020bert, naseri2021ceqe}.}
\craigc{For these baselines, we omit performances not reported in the original papers.}
%\inote{what hapenned to the other test collections. WHy are there missing rows? the reviewers went ballistic over this, we need to be clear WHY}

% Note that for Robust04, we report title-only (T) and description-only (D) queries to allow further comparisons to Neural PRF (NPRF)~\cite{li2018nprf}.

On analysing Table~\ref{tab:RQ3}, we first observe that the GenQR models significantly outperform the other \xa{generative} neural query reformulation models, which have been trained on the same training input (namely GPT2-QR, Seq2seq$_{attention}$ and Transformer). This emphasises the usefulness of using \xa{T5 and FlanT5 models} over other text-to-text approaches such as GPT2.

%. These \xiao{differences} are significant on both Robust04 (T) and (D) across both evaluation measures, while some significant differences are observed for TREC 2019 for all \xiao{neural query reformulation} approaches except GPT2-QR (for which our T5PRF TopP approach outperforms by 12\% in terms of MAP, and 10\% in terms of nDCG@10).

\looseness -1 Next, we compare \xa{the GenPRF models}
% T5PRF 
with standard PRF query expansion baselines such as RM3, Bo1 and KL. \xiaow{We observe that on \io{the} Robust04 (T) query \io{set}, \io{the} T5PRF models exhibit \io{significant} improvements over \io{the} BM25+RM3 model
% \xf{on all the compared}
for both 
metrics and significantly outperform \io{the} BM25+KL model in terms of MAP on Robust04 (D), due to the good ability of the T5 model \io{in} interpreting and reformulating the natural language queries posted in the description (D) queries.}

% In general, there are no significant differences observed here, with performances exhibited by T5PRF being broadly comparable to each of the PRF approaches. This observation demonstrates that a T5 model, when trained with examples of similar queries and passages of text giving them some context, is able to at least do as well as standard statistical PRF query expansion approaches, without modelling concepts such as IDF, and with much less pseudo-relevance information (only 2 context passages rather than 3 feedback documents). 
%is able to achieve a similar job to standard pseudo-relevance feedback models. 
% Moreover, further improvements can be attained when T5PRF is combined with neural re-rankers (as per Table~\ref{tab:RQ4}). 

\looseness -1 \xiaow{\io{Furthermore, in comparison to} DocT5query on the TREC 2019 document test query set, \io{we observe} that both \xa{T5PRF and GenPRF models}
% T5QR and T5PRF 
exhibit higher performances than the DocT5query model \xf{in terms of nDCG@10}.} 
This observation \io{demonstrates} the effectiveness of our \xa{generative reformulation models}, which do not require applying the GPU-intensive application of a T5 model to each document in the collection, while DocT5query does.

\looseness -1 \xiaoJ{When comparing with \craigj{the} CEQE models, we find that \craigj{the} T5PRF models exhibit similar performances in terms of MAP but much higher performances in terms of nDCG@20 for TREC 2019 Document ranking \craigj{query set}. For \craigj{the}  Robust title and description query sets, T5PRF models show similar performances \craigj{to the} CEQE variants on both metrics. Finally, T5PRF models exhibit slightly higher performance than \craigj{the} CEQE models on both metrics for \craigj{the} GOV2 queries.
% This indicates the T5PRF models are capable to generate useful expansion terms that possess the traits of contextualised BERT embedding based similarity to the original query terms. 
In addition, another strength of T5PRF models compared to the CEQE variants is that T5PRF models only add $\sim$10 expansion terms rather than \craigj{the} 70-100 expansion terms added by CEQE.}

% \pageenlarge{1}
Finally, we compare \craigc{with} the results reported for Neural PRF, BERT-QE. It is clear that T5PRF performs very \craigc{similarly} to Neural PRF (e.g. for MAP 0.296 vs. 0.290 on title-only queries and 0.285  vs.\ 0.280 on description-only queries). However, \io{as can be observed in Table~\ref{tab:RQ3}},  BERT-QE, which deploys three stages of BERT-Large (in comparison to our use of the comparably simpler T5-base model), exhibits a higher performance than our (simpler) T5PRF approach,  \io{which} does not employ any neural re-ranking. 
Overall, we conclude for RQ3 that T5PRF offers a promising approach for neural query expansion, which \xiaow{significantly outperforms existing statistical query expansion approaches}.

% Moreover, compared to classical query expansion approaches, it generates queries which are compatible with neural re-rankers.

%%%%%%%%%%%%%%%%%%%%%%%%%%%%NEW TABLE MONOT5%%%%%%%%%%%%%%%%%%%%%%%%%%%%%%%%%
\begin{table*}[h!]
\caption {RQ4: Performances of \xiaow{T5QR and T5PRF} using \xiaoJ{the monoT5 re-ranker. }
% \inote{check the res for gov2.genprf.monoT5. text is nan}}
% neural re-rankers. 
\craigc{\io{Notations} as in Tables~\ref{tab:RQ1} \& \ref{tab:RQ2}}.
} \vspace{-1\baselineskip}  

\centering
\def\ttestB{$\dagger$}
\resizebox{180mm}{!}{
\begin{tabular}{l cccc cccc ccc ccc ccc}
%  \hline \hline 
 \toprule
  \multirow{2}{*}{\bf{QR Models}}
  & \multicolumn{4}{c }{\bf{Document TREC 2019}} 
  & \multicolumn{4}{c }{\bf{Passage TREC 2019}} 
  &\multicolumn{3}{c }{\bf{Robust04 (T)}} 
  &\multicolumn{3}{c }{\bf{Robust04 (D)}}
  & \multicolumn{3}{c}{\bf{GOV2}}   \\
\cmidrule{2-18}

  &  MAP  & MRR & R@1k & nDCG@10
  &  MAP  & MRR & R@1k & nDCG@10   
  &  MAP   & R@1k & nDCG@10  
  &  MAP   & R@1k & nDCG@10 
  &  MAP   & R@1k & nDCG@10
  \\

\midrule

%  & & \multicolumn{16}{c}{DPH $\Rightarrow$  $ ColBERT(q^0,d) +\varphi ColBERT(q\prime, d)$ $(\varphi=0)$}\\
 
  Initial query (a) 
  &.388 &.938 &.701 &.688
  &.480 &{.857} &.752 &{.711}  
  &.269  &.708 & \bf{.481}
  &.280  &.672 & .517
  &.262 &.643 &.532

  \\ 
  BM25+RM3 (b)
&.405 &.938 &.760 &.693
&.489 &.833 &.796 &.708 
&.267 &.755 &.475
&.290 &.712 &.516
&.263 &.656 &.517

  \\ 
%   \hline
  \midrule
  
  $T5QR$  %case622
  &.394  &.938 &.753$^{a}$ &.690
  &.479  &.831 &.788$^{a}$ &.696  
  &\bf{.269}  &.759$^{a}$  &.474
  &.295  &.736 &.524
  &.269$^{b}$  &.676$^{ab}$ &.526
  \\

% $T5PRF_{FirstP}$& $\mathcal{W}_{(E+S)}$    %case622
%  &\bf{.409} &.961 &.770$^{a}$ &.696
%  &-  &- &-  &- 
%  &.268         &.757$^{a}$  &.4732
%  &.299$^{ab}$ &.743$^{ab}$ &.524
% &.270$^{ab}$ &.676$^{ab}$ &.518
%   \\
$T5PRF$    %case622
 &.403 &.926 &.776$^{a}$ &.687
 &.490 &.826 &.814$^{a}$ &.706 
 &\bf{.269} &\bf{.759}$^{a}$  &.474
 &.299$^{ab}$ &.746$^{ab}$ &.525
 &\bf{.273}$^{ab}$ &\bf{.685}$^{ab}$ &.535$^{b}$

  \\ 

 %  $T5PRF_{MaxP}$ &$\mathcal{W}_{(E+S)}$    %case622
 % &.403 &.926 &\bf{.776}$^{a}$ &.687
 %  &-  &- &-  &- 
 %  &\bf{.269} &\bf{.759}$^{a}$  &.474
 %  &\bf{.299}$^{ab}$ &\bf{.746}$^{ab}$ &\bf{.525}
 %  &\bf{.273}$^{ab}$ &\bf{.685}$^{ab}$ &\bf{.535}$^{b}$
 %  \\

% \hline
%  &  & \multicolumn{16}{c}{DPH $\Rightarrow$  $ ColBERT(q^0,d) +\varphi ColBERT(q\prime, d)$ $(\varphi=1)$}\\

   $FlanQR$ 
   &\bf{.413}$^{a}$ &\bf{.950} &\bf{.763}$^{a}$ &\bf{.699}
   &.521$^{a}$ &\bf{.908} &.845$^{a}$ &\bf{.727}
   &.265 &.756$^{a}$ &.473
   &\bf{.301}$^{ab}$ &\bf{.760}$^{ab}$ &\bf{.536}$^{ab}$
   &.271$^{a}$ &.680$^{ab}$ &\bf{.543}$^{ab}$
   \\
   % $FlanPRF_{FirstP}$  &- 
   % &&&&
   % &-  &- &-  &- 
   % &.
   % \\
   $FlanPRF$ 
   % &\bf{.456} &.828 &\bf{.841} &.697
   &.406 &.942 &.762$^{a}$ &.697
   &\bf{.530}$^{a}$  &.873  &\bf{.866}$^{a}$  &.724 
   % &.260 &.699 &.411
   &.254 &.699 &.466
   &.275 &.656 &.517
   &.241 &.597 &.524 

   \\
   % $FlanPRF_{MaxP}$ &-
   %    &&&&
   % &-  &- &-  &- 
   % \\
\bottomrule
\end{tabular}}
% }
% \end{ruledtabular}
\label{tab:RQ4:monoT5}\vspace{-0.8\baselineskip}
\end{table*}
%%%%%%%%%%%%%%%%%%%%%

%  
\pageenlarge{1} \subsection{RQ4: Integrating with Neural Re-rankers}\label{ssec:RQ4}
% \inote{rewrite this section}
% \inote{check} 
% \pageenlarge{2} 
We now investigate the effectiveness of \xa{GenQR and GenPRF models}
% T5QR and T5PRF 
when combined with \xiaoJ{a monoT5 neural \craigj{re-ranker}, in Table~\ref{tab:RQ4:monoT5}}. We integrate the reformulated queries with \craigc{the} \xiaow{monoT5 neural re-ranker} as described in Section~\ref{ssec:retrieval_setup}. 
\looseness -1 On analysing Table~\ref{tab:RQ4:monoT5}, we observe that when combined with the monoT5 re-ranker, the T5PRF models exhibit higher performances than the T5QR model, which aligns with the conclusions of RQ2. \xa{Moreover, similar to the findings of RQ2, FlanPRF leads to higher effectiveness than FlanQR on MSMARCO document and passage queries.}
We \xa{also} observe that \xa{all the generative models, except FlanPRF, significantly outperform the BM25 with monoT5 re-ranking} across all datasets in terms of Recall. This suggests that the reformulated queries generated by T5PRF are more effective in retrieving relevant documents compared to the original query or the query reformulated using BM25. \xa{For FlanPRF, we find that it excels at the MSMARCO queries but fails to produce a better reformulation for Robust and GOV2 queries.} Moreover, we find that T5PRF models show considerable improvements over both baselines in terms of MAP and Recall for the Robust04 dataset, which uses only descriptions, and the GOV2 query sets. 
We note that higher performance on Robust has been reported using monoT5 \craigj{in}~\cite{nogueira2020document}, but with an experimental setup that is not directly comparable with ours\footnote{As~\cite{nogueira2020document} makes use of {\em both} the title (T) and description (D) versions of the query, which is not a realistic setting from a user's perspective.}.
In our experiments, we use a more realistic setting where the user is expected to only enter either keyword-based queries or natural-language queries, not both.
% Overall, in response to RQ4, we conclude that T5PRF can further enhance effectiveness when combined with neural re-rankers. 
\xa{Overall, in response to RQ4, we conclude that our generative models can further enhance effectiveness when combined with neural re-rankers.}

\subsection{Discussion}
In the previous sections, we investigated the performance of all four proposed generative query reformulation methods compared to various query reformulation baselines. In this section, we discuss how the prompting-based models, FlanQR and Flan-PRF, perform compare to the fine-tuning based models, T5QR and T5PRF.
% the strengths and shortcomings of both the fine-tuning and prompting methods for implementing language models for query reformulation from various perspectives.

\pageenlarge{2} \textbf{Architecture:}
\textbf{GenQR methods} do not rely on \smmm{a set of} initial retrieved results.\footnote{\smmm{We note that although we found T5QR to be most effective when interpolated with RM3 expansion terms, the initial retrieval process for RM3 expansion can be conducted in parallel with T5QR inference, since it doe snot depend on the first-stage results.}} FlanQR has the simplest model structure as it also does not require \smmm{fine-tuning} of the model's parameters. However, FlanQR is sensitive to the input prompts to generate good query reformulations. In contrast, T5QR involves injecting the task-related knowledge into the model's parameter during fine-tuning, but once the model is trained, there is no need for prompt crafting during inference. \textbf{GenPRF methods} \smmm{depend upon} an initial round of retrieval to provide pseudo-relevant contextual information. T5PRF is fine-tuned in a weakly supervised way and can better extract useful information from the PRF contextual information. \xa{However, the only input information source for FlanPRF is prompt. As the prompt can be long, it may contain too much information for a FlanPRF model to discern which pieces of text can make a good query reformulation.}

% \inote{anything about efficiency?}
\pageenlarge{1} 
\looseness -1  \textbf{Challenges:} In fine-tuning-based methods, the primary challenge lies in constructing high-quality training data for a specific task to adapt the model's capacity to address the target task. Evidence for this can be found in the results presented in Table~\ref{tab:RQ1}. On the other hand, prompting methods do not involve changing the model's parameters and only rely on the model's existing knowledge and understanding to produce desired outputs. The key challenge for prompting methods is to identify task-related prompts that effectively unlock the large language model's learned knowledge. \xa{
Additionally, based on the qualitative study discussed in Appendix~\ref{app:case_study}, we find that the zero-shot prompting method can be challenging for the FlanPRF method in terms of identifying useful information and filtering out distracting details from the contextual information. Moreover, one might resort to few-shot prompting by providing a prompt with a few examples for generative models to learn from. However, few-shot prompting also encounters similar challenges with T5-based methods, such as constructing high-quality examples and limitations in input length.} 
Overall, based on the effectiveness results for FlanQR and FlanPRF models observed in Tables~\ref{tab:RQ2}-\ref{tab:RQ4:monoT5}, it is still promising to explore the effective way of designing query reformulation task-related prompts for the LLMs. For instance, instead of designing one prompt template and deploying it across various datasets, different prompts may be needed for different datasets. 

% \inote{what about few-shot - giving a prompt with examples and then adding the new query at the bottom?}

% \textbf{Future Work:} 

% Prompt Format:  iterative testing: validation on one task then employ the same prompts for other datasets will yield underperformed results.
% Using different prompts for the same task can help in evaluating the model's understanding of the task and its ability to generalize across various prompts.

% 1. within the GenQR framework, which doesn't rely on the initial retrieved results;

% 2. within the GenPRF framework, contextual information would be more helpful for T5PRF, which means, the additional pseudo-relevance feedback information will help the T5 model to 
% while contextual information doesn't lead to too much improvements for FLAN-PRF. Maybe because the prompting method will be sensitive to the input prompts.

% The prompting approach direct stimulates the large language model to output the desired outputs for the downstream task, without the need to modify the 
% model's parameter.

% \textbf{C}
% compare table 

% GenQR vs. RM3

% compare to other LLM query expansion, give a table includes query2doc
% \begin{table}[]
%     \centering
%     \begin{tabular}{c|c}
%          &  \\
%          & 
%     \end{tabular}
%     \caption{Caption}
%     \label{tab:my_label}
% \end{table}

\pageenlarge{2}
\section{Conclusions}\label{sec:conclusion}
% \inote{rework}
\looseness -1 In this work, we investigated neural query reformulation methods built upon the \xa{small generative neural models, such as T5 and FLAN-T5 models. In particular, we propose two possible generative query reformulation frameworks, GenQR and GenPRF. Models under the GenQR framework directly take a query as input, while models under the GenPRF framework also incorporate contextual information extracted from the pseudo-relevant feedback documents. Moreover, under each framework, we investigated both fine-tuning and direct prompting methods to leverage the learned knowledge of T5 and FLAN-T5, respectively. Extensive experiments showed that the GenQR and GenPRF models can significantly enhance effectiveness on four standard TREC test collections (when using either keyword or natural-language queries) and that significant improvements can also be observed when combined with additional neural re-rankers compared to standard PRF techniques such as RM3.}

\looseness-1 Overall, we found promise in sequence-to-sequence-based neural query reformulation models. Specifically, these can bring queries that are more precise than RM3 (benefiting nDCG@10) while also enhancing recall. Compared to competing techniques, such as docT5query, our studied models can be applied at querying time, without the need to apply expensive neural models to all documents at indexing time.
% \xf{In future work, we will continue to examine the role of further advanced weak-supervision techniques, such as query performance predictors, to aid in fine-tuning effective query reformulation models.}
% \inote{or do we change this future work?}
\xa{In future work, we will focus on advancing our understanding of how to obtain valuable pseudo-relevance feedback data effectively. \xf{In addition, we will investigate techniques for more efficient generative query reformulation. Finally,} we will delve into iterative prompting engineering for query reformulation based on the large language models.}

\bibliographystyle{ACM-Reference-Format}
\bibliography{reference}

\appendix




% Etiam c

\section{Case Study}\label{app:case_study}
% \inote{more examples for Flan version models}
Table~\ref{tab:context_example} provides (\xiaow{stemmed}) example reformulations \iadh{for} a query from \iadh{the} TREC 2019 test query set using \xiaow{the RM3~\cite{abdul2004umass},} \iadh{the} \xa{GenQR and GenPRF query reformulation models.}.

In particular, all \io{the} \xa{GenQR and GenPRF} models can generate new useful terms \craigc{that are not identified} by the RM3 model. 

\xa{For T5-based models, we observe that both T5QR and T5PRF models tend to generate more conservative terms, which present less risk for topical drift.} \xa{In addition, we observe that, in Table~\ref{tab:context_example}, T5PRF focuses on generating expansion terms that are very closely related to the definition of \textit{visceral}, such as `viscera', `body'. \io{On the other hand,} RM3 identifies terms more widely related to the body, such as `cardiac', and `fat', but which are less likely to identify more relevant documents for this query.}

\xa{For \xa{FLAN-based} models, under the GenQR framework, we find that FlanQR can generate the query reformulations consisting of natural language definitions of the \textit{visceral} from various aspects, for instance, the functionality of visceral, the further explanation of \textit{viable organ} etc. However, with the additional contextual information, some of the reformulated sentences tend to focus on \textit{abdominal}, which is provided in the contextual input, instead of the query term \textit{visceral}. Therefore, we find that FlanPRF can easily to be drifted away and might struggle to locate useful information from the potentially relevant information.}

% Thus, the query reformulations generated by the T5PRF models are capable of retrieving relevant documents that are not identified by the RM3 reformulated queries, which can be supported by the higher recall of T5PRF than RM3 in Table~\ref{tab:RQ2}.
% This observation indicates that query reformulations generated by the T5PRF models are capable of retrieving relevant documents that \io{are} not identified by \io{the} RM3 reformulated queries.
% \inote{double check!!! examples in appendix}
% \craigc{This is also apparent in the \io{example} shown in Table~\ref{tab:context_example} - T5PRF focuses on expansion terms very closely related to the definition of \textit{visceral}, such as `viscera', `body'. \io{On the other hand,} RM3 identifies terms more widely related to the body, such as `cardiac', and `fat', but which are less likely to identify more relevant documents for this query.}

% \craig{On the other hand, while the generated terms from the \xa{GenQR and GenPRF} models are more conservative, they appear to present less risk for topical drift.}}

% For FlanQR and FlanPRF models, we list 
% \inote{more analysis for FLANQR and FLANPRF}

\begin{table*}[tb]
\caption {Example reformulations of `define visceral' using \xiaow{RM3}, T5QR \& T5PRF as well as FlanQR \& FlanPRF approaches. To aid the reading, we highlight the prompts words in red and the original query in blue colour.%\inote{add importance colour}
}\vspace{-1\baselineskip}
\resizebox{180mm}{!}{
\begin{tabular}{m{180mm}}
 \toprule
 \textbf{BM25+RM3 Input Query:} \textcolor{blue}{define visceral}\\
 \textbf{BM25+RM3 Reformulated Query:}\\
 central$\wedge$0.0253 defin$\wedge$0.3167 cell$\wedge$0.0503 fat$\wedge$0.0371 obes$\wedge$0.0384 muscl$\wedge$0.077 cardiac$\wedge$0.0201 viscer$\wedge$0.3833 tissu$\wedge$0.0249 skelet$\wedge$0.0261\\
 \midrule
 {\textbf{T5QR Input Query:}} 
 \texttt{ \textcolor{red}{``refine'':}} \textcolor{blue}{define visceral}\\
  \textbf{T5QR Reformulated Query:}\\
%  viscera$\wedge$0.0621 defin$\wedge$0.0621  defin$\wedge$0.0289 viscera$\wedge$0.02896  what$\wedge$0.0196 is$\wedge$0.0196 the$\wedge$ 0.0196 viscera$\wedge$0.0196  what$\wedge$0.0153 is$\wedge$ 0.0153 a$\wedge$0.0153 viscera$\wedge$0.0153  what$\wedge$0.0073 is$\wedge$0.0073 viscera$\wedge$0.0073 \\ %62
 viscer$\wedge$0.1041 definit$\wedge$0.1041 viscera$\wedge$0.0327 defin$\wedge$0.0327  definit$\wedge$0.0253 viscer$\wedge$0.0253  viscer$\wedge$0.0135 fluid$\wedge$0.0135  viscer$\wedge$0.0098 structur$\wedge$0.0098\\ %732
    \midrule
{\textbf{T5PRF Contextualised Input Query:}} \\
% $\reprompt$: 
\texttt{ \textcolor{red}{``refine'':}} \textcolor{blue}{define visceral}.  \texttt{ \textcolor{red}{``context'':}} 
% $\coprompt$:
viscera on 5 Apr.2018. These example sentences are selected automatically from various online news sources to reflect current usage of the word `visceral'. Views expressed in the examples do not represent the opinion of Merriam-Webster or its editors. Send us feedback. The Tie Between visceral and Biology. The ``viscera'' are the internal organs of the body-especially those located in the large cavity of the trunk (e.g., the heart, liver, and intestines). The word viscera comes from Latin, in which it has essentially the same meaning.Something ``visceral'' has to do with the viscera. In a more figurative sense, something ``visceral'' is felt ``deep down''. Even in the early years of its use, ``visceral'' often referred to things emotional rather than physiological. For example, in 1640, an English bishop named Edward Reynolds wrote, ``Love is of all other the inmost''.\\
 \textbf{T5PRF Reformulated Query:}\\
 viscera$\wedge$0.1596  viscer$\wedge$0.1556  viscera$\wedge$0.0592 bodi$\wedge$0.0592  viscer$\wedge$0.0379 fat$\wedge$0.0379 definit$\wedge$0.0243 viscer$\wedge$0.0243  \\
\midrule
\textbf{FlanQR Input Query:} \texttt{ \textcolor{red}{Improve the search effectiveness by suggesting expansion terms for the query:}} \textcolor{blue}{define visceral}\\
 \textbf{FlanQR Reformulated Query:} \\
Viable organ means the possibility that it could be dead if not harvested and the normal, healthy human being is capable of dead Organs, substances in organ meat and whole part of digestion\\
Viscera (and viscerally) Able:-able; competent, eager to serve\\
Visceral or viscular can mean the inner layer, which includes any cell wall-containing tissue lining cells in organs\\
Visceral art refers to the expressive medium of artists, drawing on recognizable visceral materials: bones, organs, muscle tissue, soft tissue, cells, chemicals, fat and saliva\\
Visco-elastic properties, properties of visceral organs, measure of stiffnesss\\
 
\midrule
\textbf{FlanPRF Input Query:} \texttt{ \textcolor{red}{Improve the search effectiveness by suggesting expansion terms for the query:}} \textcolor{blue}{define visceral}, \texttt{\textcolor{red}{based on the given context information:}} Abdominal fat can be visceral or subcutaneous. Visceral adipose tissue (VAT) is fat that is around the organs inside the abdomen. Subcutaneous adipose tisue (SAT) is fat under your skin (love handles). With visceral fat your stomach wall is pushed out from inside. Your stomach muscles can sometimes be quite defined, but your stomach will still be extended. isceral adipose tissue (VAT) is fat that is around the organs inside the abdomen. Subcutaneous adipose tisue (SAT) is fat under your skin (love handles). With visceral fat your stomach wall is pushed out from inside. Your stomach muscles can sometimes be quite defined, but your stomach will still be extended. \\
\textbf{FlanPRF Reformulated Query:} \\

Visceral adipose tissue (VAT) is fat that is around the organs inside the abdomen.\\
Visceral (adjective) Abdominal fat can be visceral or subcutaneous.\\
Abdominal fat can be visceral or subcutaneous. Visceral adipose tissue (VAT) is fat that is around the organs inside the abdomen.\\
Visceral adipose tissue (VAT) Visceral adipose tissue (VAT) is fat that is around the organs inside the abdomen.\\
Abdominal fat can be visceral or subcutaneous. Visceral adipose tissue (VAT) is fat that is around the organs inside the abdomen. \\

\bottomrule
\end{tabular}}\label{tab:context_example}
  
\end{table*}

\section{Fine-tuning process of T5QR and T5PRF}\label{app:fine_tune_process}
T5~\cite{raffel2020exploring} is a \xiao{large pretrained} Text-To-Text-Transfer-Trans\-former model. 
% pretrained on \iadh{the} C4 dataset\inote{never heard of it, why is C4 important?}.
\iadh{Within} such a sequence-to-sequence framework, all the tasks can be \iadh{cast in a} Text-to-Text format. After pre-training, the pretrained knowledge \iadh{is} stored as the parameters of a T5 model, denoted as $T5(\bar{\theta})$. 
The T5~\cite{raffel2020exploring} model \craig{consists of separate stacks of encoding and decoding layers: each encoding layer contains a self-attention layer and a feed-forward layer; while each decoding layer comprises a self-attention layer, \iadh{an} encoder-decoder attention layer and a feed-forward layer.}
The encoder takes  a source sequence $X=(x_1, x_2, ...,x_n)$ with a length of \xiaow{$n$} terms and the model is trained to produce the corresponding target sequence ${Y}=({y_1}, {y_2}, ..., {y_m})$ with a length of \xiaow{$m$} terms.  
\craig{The ultimate goal when training a T5 model is, given a set of training pairs $ \mathcal{M} = \{\langle X,Y \rangle \}$, 
% is 
for each pair  $\langle X,Y \rangle$ to maximise the posterior for \iadh{a} target output $Y$ given an input sequence $X$.} 
The T5 decoder produces one token at each time step, thus at each time step during training, the maximum-likelihood \craig{objective} function is: 
\begin{equation}\label{eqnProbT5}
\max P_{T5}(Y \mid X)=\max \prod_{t=1}^{M} P_{T5}(y_{t} \mid y_{1:t-1}, X,\bar{\theta} )
\end{equation}
\looseness -1 where $y_{1:t-1}=(y_{1}, \ldots, y_{t-1})$ \craig{are} the tokens generated in  \iadh{the} previous steps. At each time step $t$, T5 \iadh{maximises} the conditional probability in Equation~\eqref{eqnProbT5} by minimising the negative log probability as the prediction loss, \craigc{also known as} the  Cross-Entropy (CE) loss:
% \begin{equation}
%  \mathcal{L}_{T5}\left(y_{t}\right)=- \sum_{k=1}^{K}\sum_{i=1}^{C} {\tau_{i}}^k \log (P_{T5} (c_{i} \mid y_{1:t-1}, X^k,\bar{\theta}))
% \end{equation}
% 
\begin{equation}~\label{T5:loss}
 \mathcal{L}_{T5}\left(y_{t}\right)=- \sum_{i=1}^{C} {\tau_{i}} \cdot \log (P_{T5} (c_{i} \mid y_{1:t-1}, X,\bar{\theta}))
\end{equation}
% \inote{check this equation,-xiao}
% \begin{equation}~\label{T5:loss}
%  \mathcal{L}_{T5}\left(y_{t}\right)=- \sum_{i=1}^{M} {y_{m}} \cdot \log (P_{T5} (c_{i} \mid y_{1:t-1}, X,\bar{\theta}))
% \end{equation}
% \inote{here, we should not use batch loss, for X, Y we just give as one example to illustrate one pass }
% \inote{$\tau_i$ is act as a function of $y_t$, it's not the target label, its the actual output class, we train the model want to make $\tau_i$ equals to the target label. for instance, in one pass, actually generated $[0,0,1]$, target output could be the first word $[1,0,0]$} 
\looseness -1 where: $c_1,c_2,...,c_C$ \iadh{are} the search space, i.e.\ the classes for the \io{decoder} -- indeed, for a text generation task, $C$ is equal to the length of the output vocabulary of the model;
% $K$ is the number of training samples in the batch. 
\craig{$\tau_{i}$ is an indicator variable equal to 1 when $y_t=c_i$ and 0 otherwise; and}
% is the target label for class $c_i$.
% \inote{shouldnt this refer back to Y?}. 
$P_{T5}(c_i|\cdot)$ is the predicted probability produced by the soft-max
% \inote{so soft-max comes from nowhere} 
layer of \iadh{the} decoder. 
% , as shown in Figure~\ref{fig2}. %\todo[craig]{we need a new version of the figure back}
% \xiao{\craig{Assuming that} the term with the highest probability is selected as the current time step’s output term, which is possibly not the ground truth term. 
The loss in Equation~\eqref{T5:loss} is backpropagated through the network to tune the model’s parameters, s.t.\ the model's output is closer to the target. 

 \sm{T5 generates text \textit{auto-regressively}; that is, the probability of a generated sequence is calculated as the product of the probability of each token in the sequence.}
% \craigc{At} test time, T5 employs the auto-regressive language generation mechanism, \craigc{which is based on the assumption that the joint probability of a sequence can be obtained from the product of the conditional probabilities obtained from predicting each word in the sequence}.
%the probability distribution of a sequence can be \craigc{decomposed} into \craigc{a product of the conditional probabilities for predicting the next word}. 
\xiaow{\craigc{By applying} a fine-tuned T5QR model, $T5QR(\hat{\theta})$, the \craigc{joint likelihood} of the query reformulation is:} %$P(q^r_1, q^r_2,..., q^r_{|q^r|})$ is:}
\begin{equation}
    P\left(q^r_1, q^r_2,..., q^r_{|q^r|}\right) = \prod_{t=1}^{|q^r|} P_{T5QR}(q^r_t|q^r_{1:t-1},q^0,\hat{\theta})\label{equ:weight}
\end{equation}

% %%%%%%%%%commented out
\looseness -1 \craigc{\io{Hence}, the final output of $\mathcal{P}_{T5QR}(q^0)$ is \craigc{obtained by combining the output sequences of $N$ applications of the T5 model, and weighting the terms from each sequence by the joint likelihood of the sequence, as follows:}}
%as the weighted combination of terms:}
\begin{equation}
    \mathcal{P}_{T5QR}(q^0) = w_{q^{r1}}\cdot \left[ q^{r1}_1,... q^{r1}_{|q^{r1}|}\right]+...+ w_{q^{rN}}\cdot \left[ q^{rN}_1,... q^{rN}_{|q^{rN}|}\right],\label{eqn:PT5QR}
\end{equation}
where $w_{q^r}$ denotes the joint likelihood of reformulation (or paraphrase of the input query) $q^r$ -- i.e.\ $w_{q^r} = P\left(q^r_1, q^r_2,..., q^r_{|q^r|}\right)$ -- and $N$ is the number of predicted output sequences used to form a query reformulation in response to the original query $q^0$. By generating and combining $N$ paraphrases of the original query generated by T5, important terms are more likely to receive higher weights -- indeed, a similar repeated application of T5 is used by docT5query~\cite{nogueira2019doc2query}.
\xa{The fine-tuning process for T5PRF is similar to T5QR, the only difference is the input template. 
The final reformulated query for T5PRF is obtained by combining the weighted combinations of $N$ invocations of Equation~\eqref{CAequ}, in a similar manner as for T5QR in Equation~\eqref{eqn:PT5QR}.}
% \inote{check this eqn nbr}.
% However, while \io{the user's} input queries may be short and \io{may not} provide sufficient evidence to interpret the meaning of the input query, in the next section we show how T5QR can be formulated as \iadh{a} pseudo-relevance feedback mechanism, following Equation~\eqref{eq:PRF}.

\section{Hyperparameter study}\label{app:hyper_param}
% \subsection{Fine-tuning }

% \subsection{RQ5: Hyperparameter Study}\label{ssec:RQ5}

% the hyperparameters in our proposed T5PRF models. 
% \inote{move hyperparameter study to appendix}
% \inote{rework here}
Finally, we aim to determine the \io{effect} of \io{the} hyperparameters in our T5-based reformulation models. 
% \noindent \textbf{RQ5:} What is the impact of the number of top $M$ passages in T5PRF, \xiaow{the number of paraphrases $N$ to form the query reformulations \craigc{as well as the relative weighting of reformulations obtained from RM3 and T5-based reformulation models}?}

\looseness -1 
% \xiaow{We address RQ5 \craigc{by analysing the importance of the hyperparameters of T5QR and T5PRF}.
We first consider the impact of the number of selected passages, $M$, used as the contextualised input to \xiao{T5PRF}. Recall that the maximum input to \xiaow{the pretrained} T5 model, $X$, is limited to 512 tokens; for this reason, and to ensure \io{sufficient} space for the initial query and the prompt tokens, $M$ cannot exceed 3 passages, given that we use passages of 128 tokens (see Section~\ref{ssec:T5_setup}).

\begin{figure*}[tb]
\centering
\begin{subfigure}[b]{0.43\textwidth}\centering
\includegraphics[width=\textwidth]{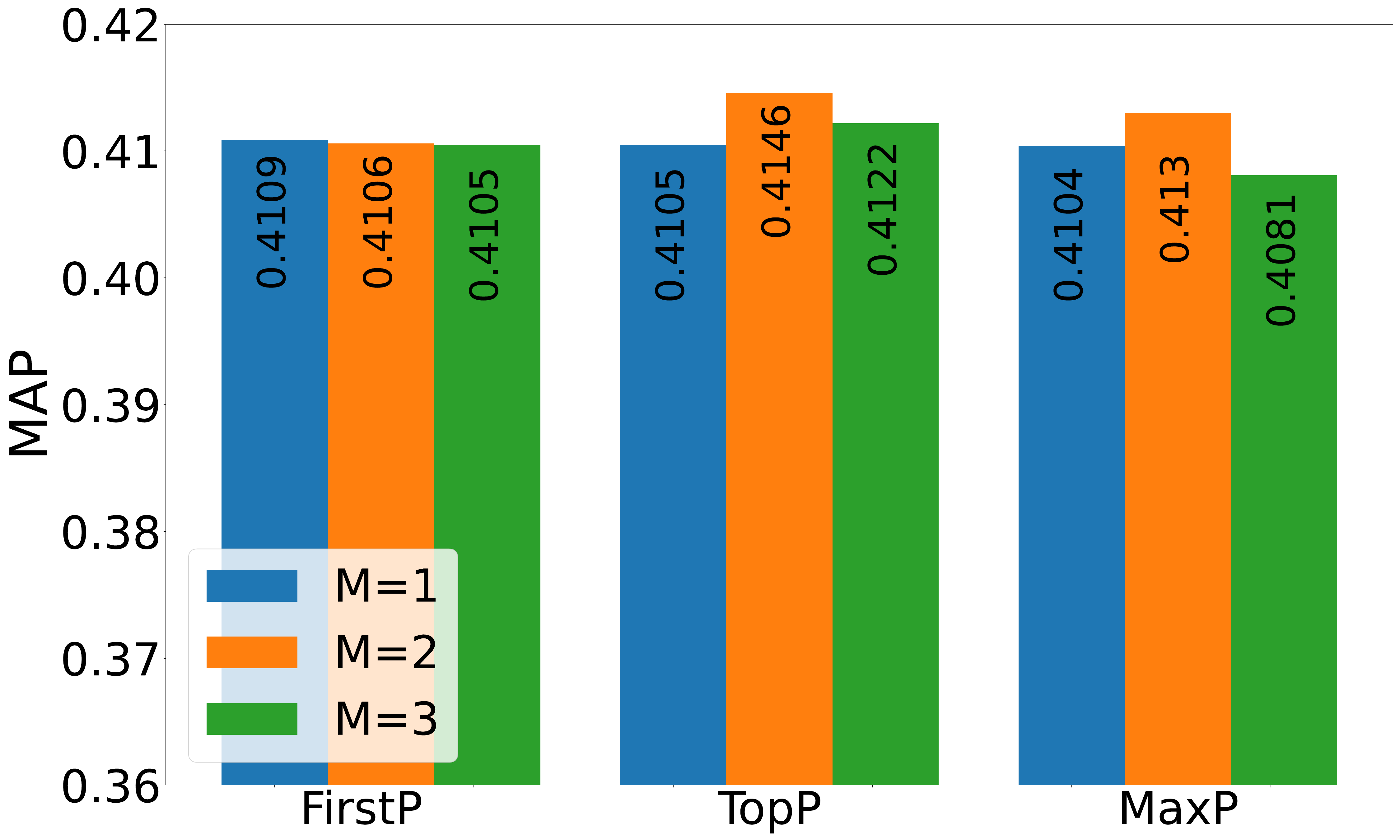}
\end{subfigure}
\begin{subfigure}[b]{0.43\textwidth}\centering
\includegraphics[width=\textwidth]{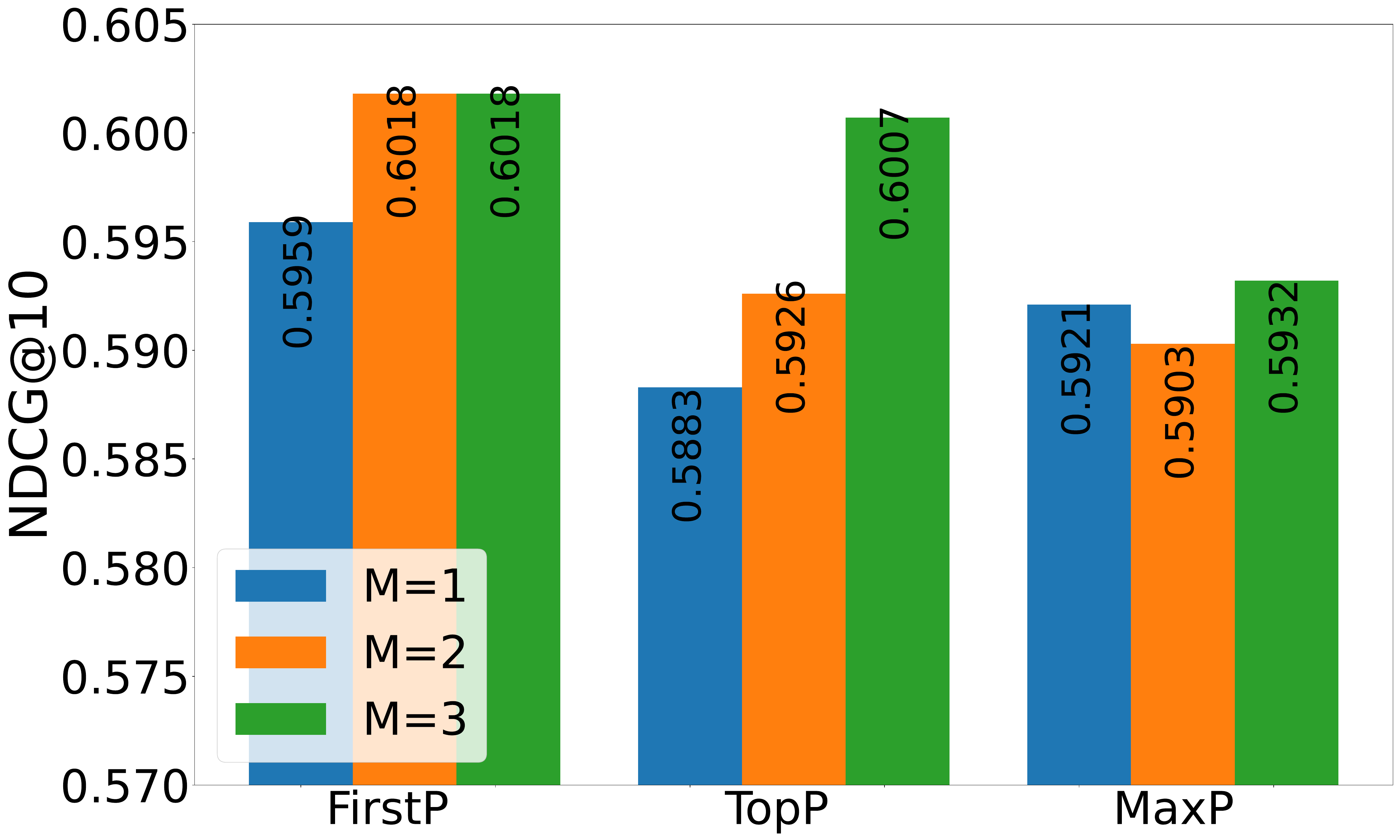}
\end{subfigure}
\caption{Impact of the number of passages $M$ in T5PRF. Different coloured bars are shown for the three different contextual passage selectors FirstP, TopP and MaxP.}
\label{fig:M}
\end{figure*}

\looseness -1 \craigc{Figure~\ref{fig:M} presents the MAP and nDCG@10 scores of the T5PRF models using different $M$ \io{values} on \io{the} TREC2019 document query set. We observe that while nDCG@10 is improved with more passages (i.e. as $M$ increases), MAP tends to be degraded. We postulate that when more passages \io{are} selected as an input, \io{it is more likely} that the T5 model will generate a few off-topic terms (i.e. topic drift), which will have low \io{weights} but can negatively affect MAP; On the other hand, with more passages, the very important terms for the most highly relevant passages are more easily identified and emphasised in the resulting query reformulation, resulting in improved nDCG@10.}

\begin{figure*}[tb]
\centering
\begin{subfigure}[b]{0.45\textwidth}\centering
\includegraphics[width=\textwidth]{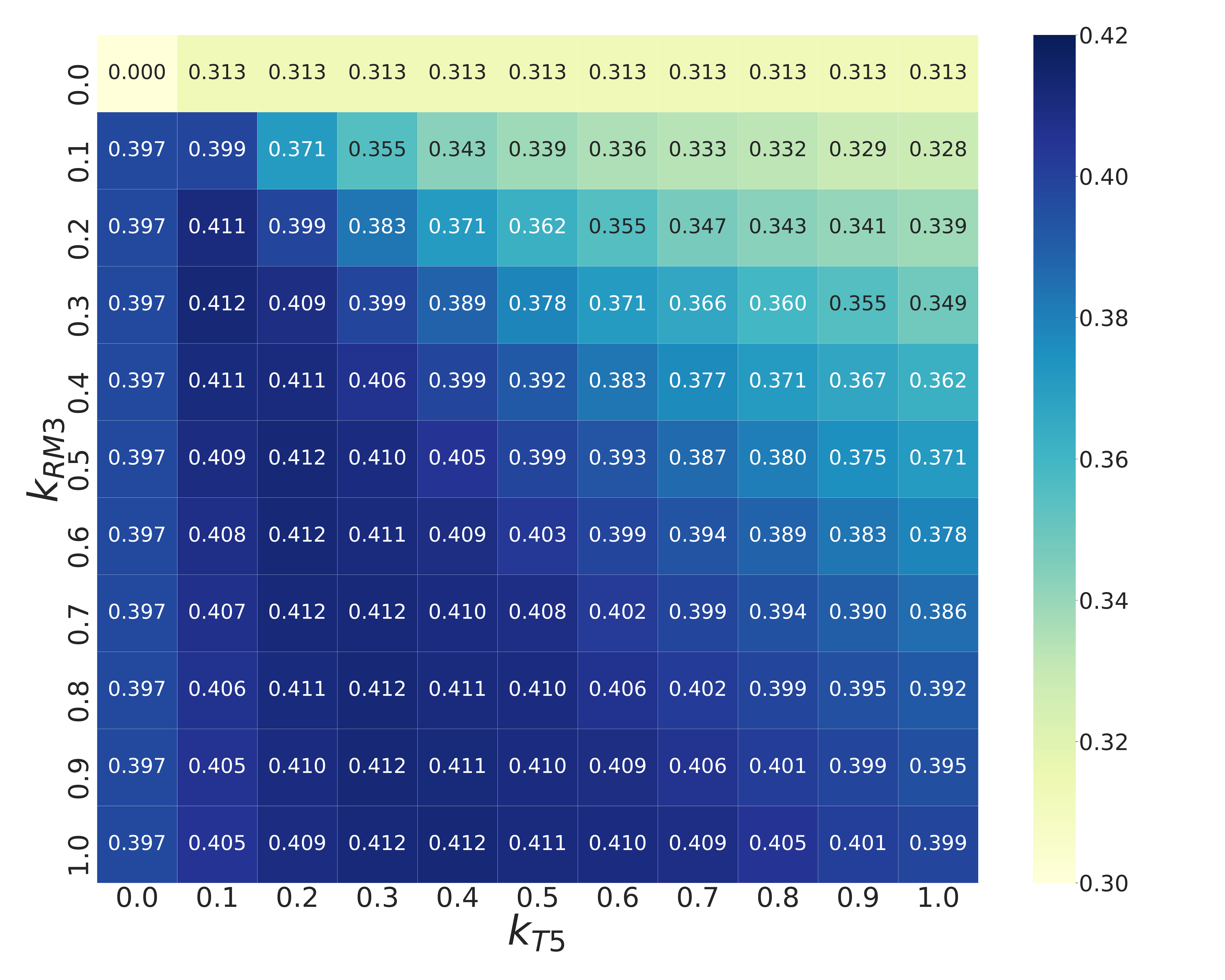}
% \caption{Query \& doc. embeddings.}\label{fig:pca:query}
\end{subfigure}
\begin{subfigure}[b]{0.45\textwidth}\centering
\includegraphics[width=\textwidth]{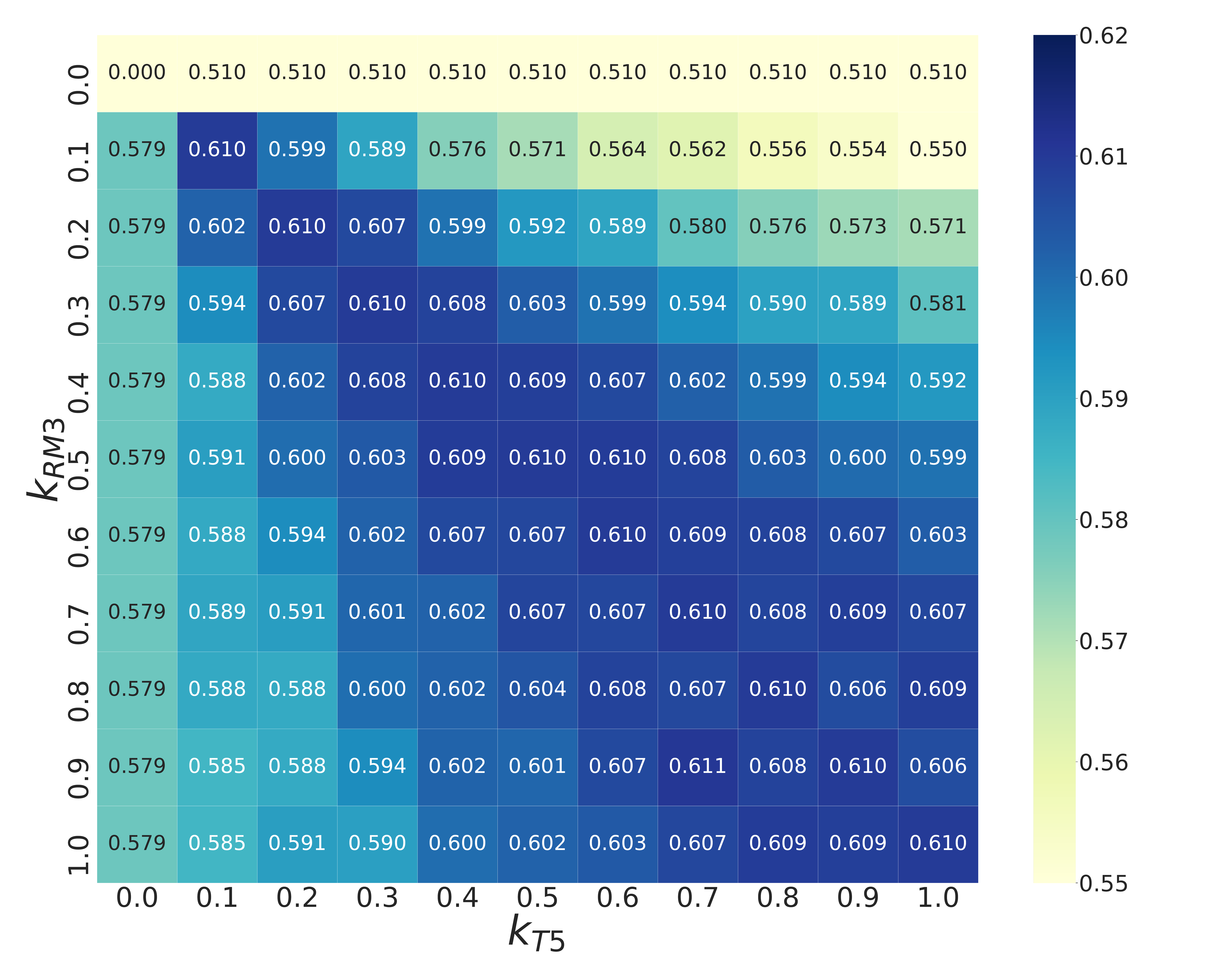}
% \caption{Cluster centroids, $\clusters=24$.}\label{fig:pca:doc}
\end{subfigure}
\caption{Impact of the mixing \io{parameters} $k_{RM3}$ and $k_{T5}$.}
\label{fig:weights}
\end{figure*}

\xiaow{Next we investigate the impact of the $k_{RM3}$ and $k_{T5}$ parameters \xiaow{introduced in Section~\ref{ssec:retrieval_setup}.}}
% , which control the \io{weights} of the query reformulations generated by RM3 and \io{the} \xiaow{T5PRF model.}}
% T5-based \io{models}, respectively.} 
\craigc{Figure}~\ref{fig:weights} \io{shows a} heatmap depicting the performance of a T5PRF model using different values of $k_{T5}$ (x-axis) and $k_{RM3}$(y-axis) on the TREC 2019 Document task, for (a) MAP and (b) nDCG@10. 
%. The cell of the first sub-figure denotes the value of MAP score using the specific combination of the two parameters while cell of the second-figure denotes the value of nDCG@10 score, the darker the cell, the higher the value. 
Interestingly, we note differences in the overall patterns between MAP and nDCG@10. For both metrics, the highest values are generally on or near to the diagonal; while effectiveness drops off for low values of $k_{T5}$. \io{This} is more marked for nDCG@10 than MAP. \craigc{We therefore conclude that when the T5-generated query reformulations are combined with RM3, there is more impact to the top of the ranking, as quantified by the larger improvements in nDCG@10 compared to MAP.}

\begin{figure*}
\centering
\begin{subfigure}[b]{0.49\textwidth}\centering
\includegraphics[width=\linewidth]{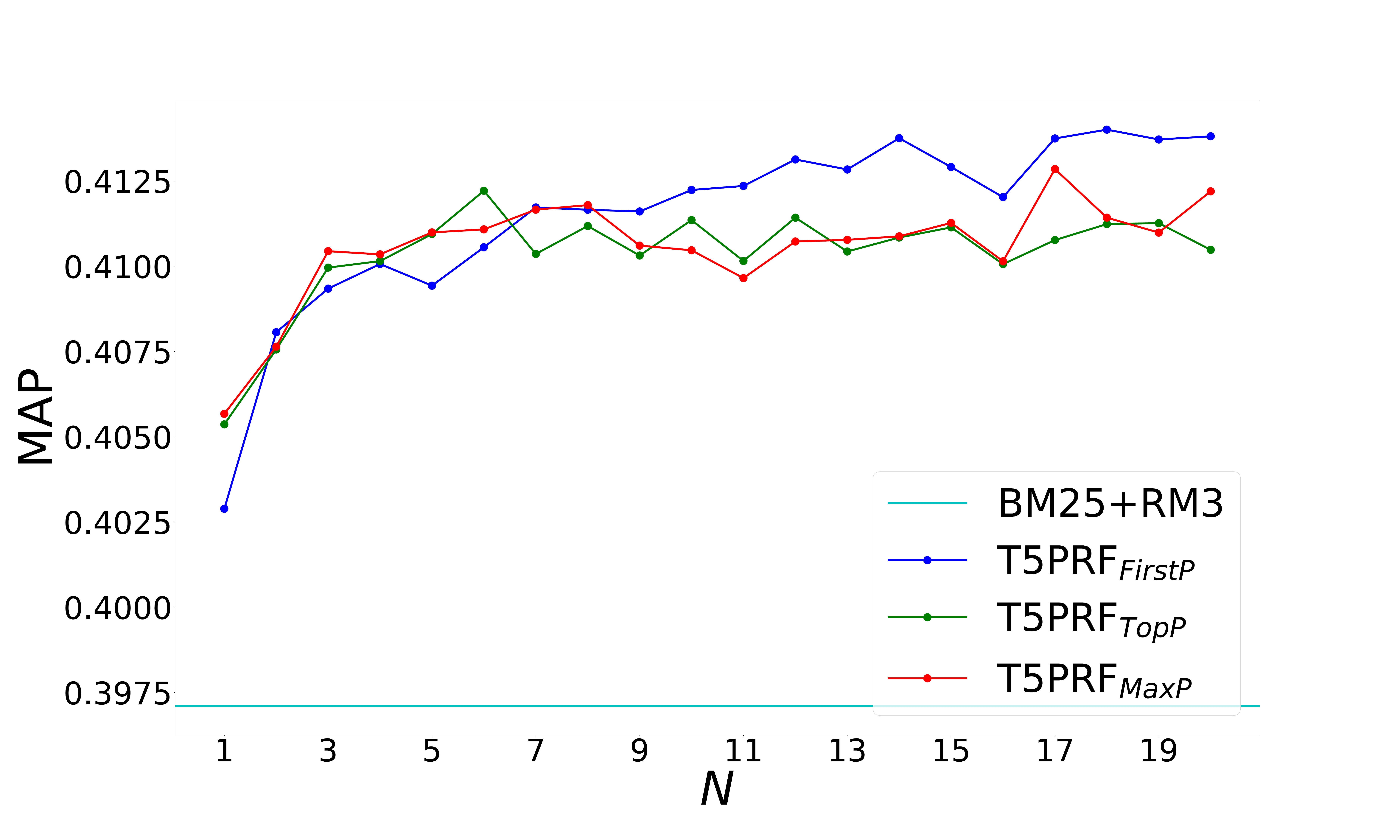}
% \caption{Query \& doc. embeddings.}\label{fig:pca:query}
\end{subfigure}
\begin{subfigure}[b]{0.49\textwidth}\centering
\includegraphics[width=\linewidth]{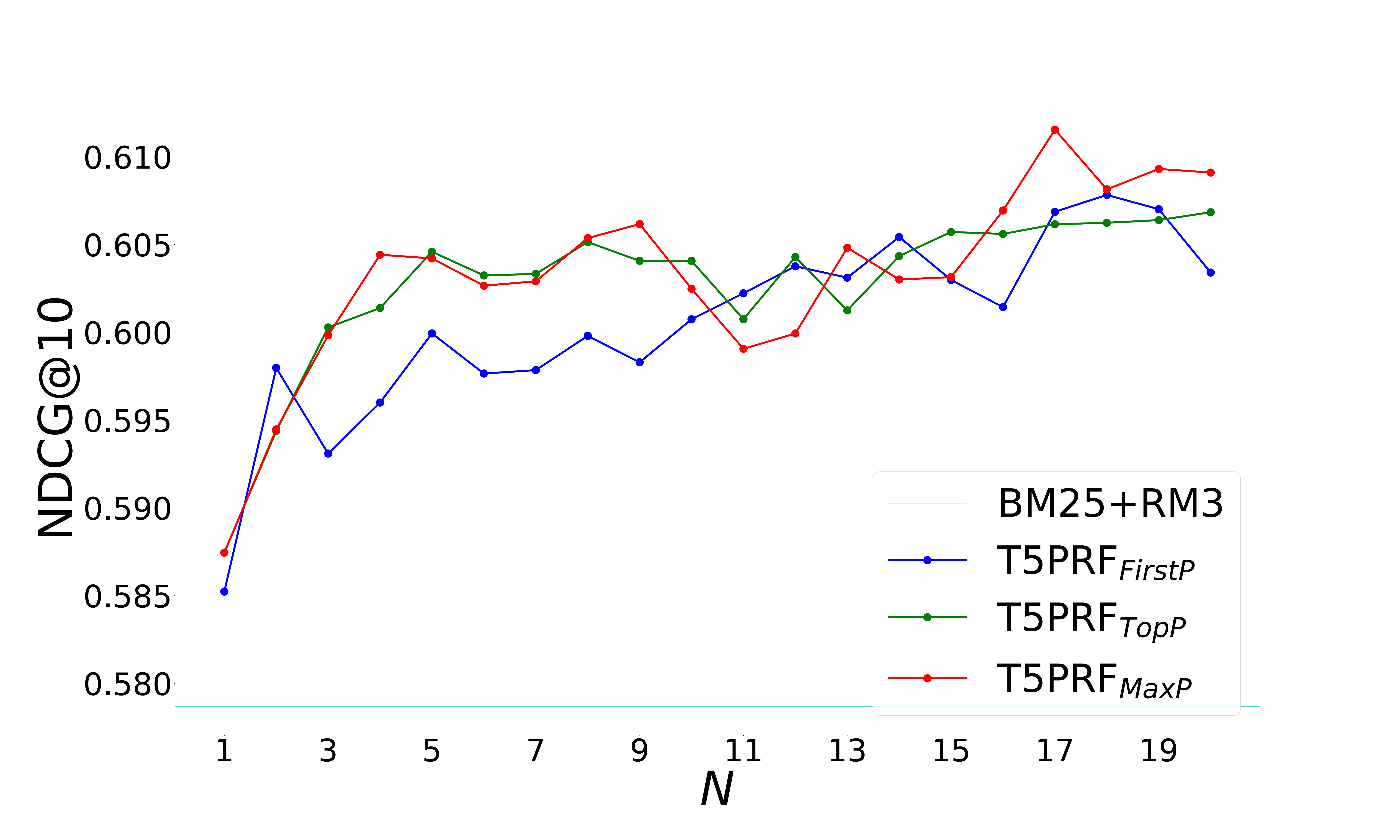}
% \caption{Cluster centroids, $\clusters=24$.}\label{fig:pca:doc}
\end{subfigure}
\caption{ Impact of the number of \xiaow{paraphrases} $N$ obtained from T5 to form a query reformulation.}
\label{fig:NRank}
\end{figure*}

\xiaow{Finally, we \craigc{examine} the impact of the number of} paraphrases $N$ \xiaow{selected \craigc{from a T5-based model} to \io{construct} the reformulated query. We take a trained model of T5PRF using the $W_{(E+S)}$ filter as an example. Figure~\ref{fig:NRank} illustrates the impact \io{on} \io{the} MAP and nDCG@10 \io{scores} \io{when} varying $N=[1,20]$. We \craigc{observe} that, in general, the higher the number of paraphrases to be included in the query reformulation, the higher \io{ the values that} MAP or nDCG@10 \io{are} likely to achieve, although with \craigc{a degree of variance}. In particular, all tested $N$ \io{values} markedly exceed the corresponding performance of BM25+RM3. Note that while we only \io{chose} $N=5$ \io{in reporting the experiments} in this paper \io{in order} to facilitate faster retrieval, \io{larger} $N$ \io{values} usually lead to \io{a} higher effectiveness.}

\end{document}